\documentclass{aa}
\usepackage{graphicx}
\usepackage{subcaption}
\usepackage{enumitem}
\usepackage{amsmath}
\usepackage{multicol}
\usepackage{txfonts}
\usepackage[colorlinks=True,pdfborder={0 0 0}, linkcolor=blue,citecolor=blue,breaklinks=true]{hyperref}

\begin{document}

   \title{Formation of low mass protostars and their circumstellar disks}

   \author{A. Ahmad
          \inst{1}
          \and
          M. González\inst{1}
          \and
          P. Hennebelle\inst{2}
          \and
          B. Commerçon\inst{3}
          }

   \institute{Université Paris Cité, Université Paris-Saclay, CEA, CNRS, AIM, F-91191, Gif-sur-Yvette, France
             \and
             Université Paris-Saclay, Université Paris Cité, CEA, CNRS, AIM, 91191, Gif-sur-Yvette, France
             \and
             Univ Lyon, Ens de Lyon, Univ Lyon 1, CNRS, Centre de Recherche Astrophysique de Lyon UMR5574, 69007, Lyon, France
             }

   \date{Received 25/01/2024; accepted 21/04/2024}

 
  \abstract
  {Understanding circumstellar disks is of prime importance in astrophysics, however, their birth process remains poorly constrained due to observational and numerical challenges. Recent numerical works have shown that the small-scale physics, often wrapped into a sub-grid model, play a crucial role in disk formation and evolution. This calls for a combined approach in which both the protostar and circumstellar disk are studied in concert.}
   {We aim to elucidate the small scale physics and constrain sub-grid parameters commonly chosen in the literature by resolving the star-disk interaction.}
   {We carry out a set of very high resolution 3D radiative-hydrodynamics simulations that self-consistently describe the collapse of a turbulent dense molecular cloud core to stellar densities. We study the birth of the protostar, the circumstellar disk, and its early evolution ($<6\ \mathrm{yr}$ after protostellar formation).}
   {Following the second gravitational collapse, the nascent protostar quickly reaches breakup velocity and sheds its surface material, thus forming a hot ($\sim 10^{3}\ \mathrm{K}$), dense, and highly flared circumstellar disk. The protostar is embedded within the disk, such that material can flow without crossing any shock fronts. The circumstellar disk mass quickly exceeds that of the protostar, and its kinematics are dominated by self-gravity. Accretion onto the disk is highly anisotropic, and accretion onto the protostar mainly occurs through material that slides on the disk surface. The polar mass flux is negligible in comparison. The radiative behavior also displays a strong anisotropy, as the polar accretion shock is shown to be supercritical whereas its equatorial counterpart is subcritical. We also find a remarkable convergence of our results with respect to initial conditions.}
   {These results reveal the structure and kinematics in the smallest spatial scales relevant to protostellar and circumstellar disk evolution. They can be used to describe accretion onto regions commonly described by sub-grid models in simulations studying larger scale physics.}


   \maketitle
%

\section{Introduction}
Circumstellar disks form as a result of the conservation of angular momentum during the collapse of gravitationaly unstable pre-stellar cloud cores. Understanding the formation of these disks and their subsequent evolution is of fundamental importance in astrophysics, as they are the birthplace of planets. This task is however heavily impeded by numerous challenges in observing star forming regions, as the dense infalling envelope obscures the nascent disk during the class 0 phase. As such, most observational constraints come from more evolved class I and II disks. From a theoretical standpoint, the prohibitive time-stepping constraints in numerical simulations has made it nearly impossible to self-consistently describe the evolution of a newly formed circumstellar disk over a sufficiently large timescale to compare it with observations. In order to circumvent these constraints, theorists have abandoned the description of the innermost regions ($< 1\ \mathrm{AU}$) and instead use sink particles \citep{Bate_1995, bleuler_2014} onto which sub-grid physics are encoded. These particles interact with the surrounding gas through self-gravity and accretion, as well as radiative and mechanical feedback effects such as outflows and stellar winds. The parameters of these particles, such as their effective radius and accretion thresholds, have largely been chosen on the grounds of educated guesswork. Although necessary to study the global evolution of the disk, reducing the inner regions (which contain the protostar) to a sub-grid model can produce nonphysical results, especially when much of the actual sub-grid physics that has been encoded remains poorly constrained.
\\
In this respect, \cite{machida_2014} investigated the effects of sink parameters on the formation of circumstellar disks. They found that the choice of the sink radius and its accretion threshold can, in conjunction with the physical model employed, dictate the formation and evolution of a circumstellar disk. \cite{vorobyov_2019} lead a similar study, this time focusing on the mass transport rate from the sink cell to the protostellar surface. They found that simulations with a slower mass transport rate would form more massive disks, and the accretion rate onto the protostar displayed more episodic behavior.
\\
Finally, \cite{hennebelle_disks} studied the influence of the sink accretion threshold on the global evolution of the disk. They found that while the mass contained within the sink is insensitive to this parameter, the disk radius and mass on the other hand exhibit a strong sensitivity to it. Indeed, they found that the disk mass increases significantly at higher accretion thresholds.
\\
\\
From these empirical studies, it has become clear that a deeper understanding of the disk's inner boundary is a necessary endeavour to pursue in order to understand the global disk evolution. Of course, this is not the only field of application of sink particles. Much larger scale simulations that seek to provide a protoplanetary disk or stellar population synthesis by modeling the collapse of an entire molecular cloud such as those of \cite{bate_2012, bate_2018, hennebelle_2020, lebreuilly_2021, grudic_2022, lebreuilly_2023, lebreuilly_2023b} also rely on sink particles for their inclusion of small-scale physics. This also applies to simulations studying the high-mass regime (e.g. \citealp{krumholz_2009, kuiper_2010, mignon_2021b, mignon_2021a, commercon_2022, mignon_2023, andre_2023}). However, understanding the inner boundaries of circumstellar disks requires a self-consistent description of the inner regions, in which one must model the formation of the protostar following a second gravitational collapse triggered by the dissociation of $\mathrm{H}_{2}$ molecules and its subsequent interaction with the disk. Although \cite{larson1969} had pioneered second collapse calculations in spherical symmetry, the field has since developed ever more robust codes to tackle the problem in the three dimensions necessary to describe the formation of a circumstellar disk, all the while including ever more complex physics such as radiative transfer (e.g, \citealp{whitehouse_2006, bate_2010, Bate_2011, ahmad_2023}) and magnetic fields under the ideal and non-ideal approximations (\citealp{machida_2006, machida_2007, machida_2008, machida_2011, machida_2011b, tomida_2013, tomida_2015, Bate_2014, tsukamoto_2015, wurster_2018, vaytet_2018, machida_2019, wurster_2020, wurster_2021, wurster_2022}).  Although the latest studies struggle to integrate across large timescales due to stringent time-stepping constraints, an important result they've shown is that the higher density gas ($\rho > 10^{-10}\ \mathrm{g}\ \mathrm{cm}^{-3}$) is poorly magnetized due to magnetic resistivities, thus placing the magnetic pressure orders of magnitude below the thermal pressure. As such, in addition to greatly alleviating numerical constraints, omitting magnetic fields allows one to describe the inner sub-AU region with reasonably high fidelity prior to the birth of a stellar magnetic field through a dynamo process. Additionally, these studies have not studied in depth the interaction between the nascent protostar and its surrounding disk, mostly due to the prohibitive time-stepping. Nevertheless, they have offered valuable insight regarding the system's structure following the second collapse phase. Indeed, they seem to indicate that the inner regions are characterized by a density plateau in the innermost region ($<10^{-2}$ AU), which then transitions towards a power-law distribution \citep{saigo_2008, machida_2011b, tsukamoto_2015, vaytet_2018}.
\\
\\
In this paper, we investigate the inner boundaries of newly-formed circumstellar disks using high resolution 3D radiation-hydrodynamics (RHD) calculations of the collapse of a dense molecular cloud core to protostellar densities under the gray flux limited diffusion approximation (FLD). To this end, we model the collapse of the pre-stellar core, the formation of a first hydrostatic Larson core, the second gravitational collapse triggered by the dissociation of $\mathrm{H}_{2}$ molecules, and the subsequent early evolution of the inner regions. Particular attention is given to the interaction between the nascent protostar and the disk and how such a process evolves over time.

\section{Model} \label{section:model}

We carried out our simulations using the {\ttfamily RAMSES} \citep{teyssier_2002} adaptive mesh refinement (AMR) code and the same setup as that employed in \cite{ahmad_2023}, with a notable difference being the presence of angular momentum in the system through the inclusion of an initially turbulent velocity vector field parameterized by a turbulent mach number $M_{\mathrm{a}}$. FLD was implemented in the code by \cite{commercon_2011, commercon_2014} and \cite{gonzalez_2015}. We have used the equation of state of \cite{saumon_1995}, and the opacity tables of \cite{semenov_2003}, \cite{ferguson_2005}, and \cite{badnell_2005}, which were pieced together by \cite{vaytet_2013}. The initial conditions consist of a uniform density sphere of mass $M_{0}=1\ \mathrm{M_{\odot}}$, initial temperature $T_{0}=10\ \mathrm{K}$, and a radius of $R_{0}=2.465\times 10^{3}\ \mathrm{AU}$, thus yielding a ratio of thermal to gravitational energy of $\alpha=0.25$. We present four runs in the main body of this paper where the initial amount of turbulence varies: $M_{\mathrm{a}}=0.2$ for run G1, 0.4 for G2, 0.8 for G3, and 1 for G4. Although $M_{\mathrm{a}}$ varies, the turbulent seed does not. This means that run G4 has five times as much initial angular momentum as G1. For comparative purposes, we also run two additional simulations labeled G5 and G6. G5 possesses the same parameters as run G2 but also includes solid body rotation in the initial cloud core whose ratio of rotational to gravitational energy $\beta_{\mathrm{rot}}=10^{-2}$. Run G6 also contains the same parameters as G2, but has a higher $\alpha$ of 0.5.
\\
We used the same refinement strategy as in \cite{ahmad_2023}, however, since angular momentum is present in the system, the simulations do not require a resolution as stringent as their spherically symmetrical counterpart as the properties of the protostar, such as its central density and radius, are easier to resolve. Thus, we lower the maximum refinement level $\ell_{\mathrm{max}}$ to 25 in order to alleviate our time-stepping constraints, thus yielding a spatial resolution of $\Delta x=2.9\times10^{-4}\ \mathrm{AU}$ at the finest level (the coarse grid is $64^{3}$ cells, with $\ell_{\mathrm{min}}=6$). Nevertheless, the stringent refinement criterion yields some of the best resolved disks in the literature. Indeed, the circumstellar disks in our simulations have $[95-2.7\times 10^{3}]$ cells per Jeans length, with $\sim 10^{7}$ cells within their volume.

\section{Results} \label{section:results}
Our simulations cover the initial isothermal contraction of the cloud, the birth of the first Larson core, the second gravitational collapse, and the subsequent evolution of the star-disk system. The physics at large scales (i.e., from the cloud core to the first Larson core) has been thoroughly discussed in the literature, and as such will be briefly covered in our paper. Below, we focus our study on the behavior of the system in the innermost regions that contain the protostar.

\subsection{The dynamical range}
\begin{figure*}
    \centering
    \includegraphics[width=.9\textwidth]{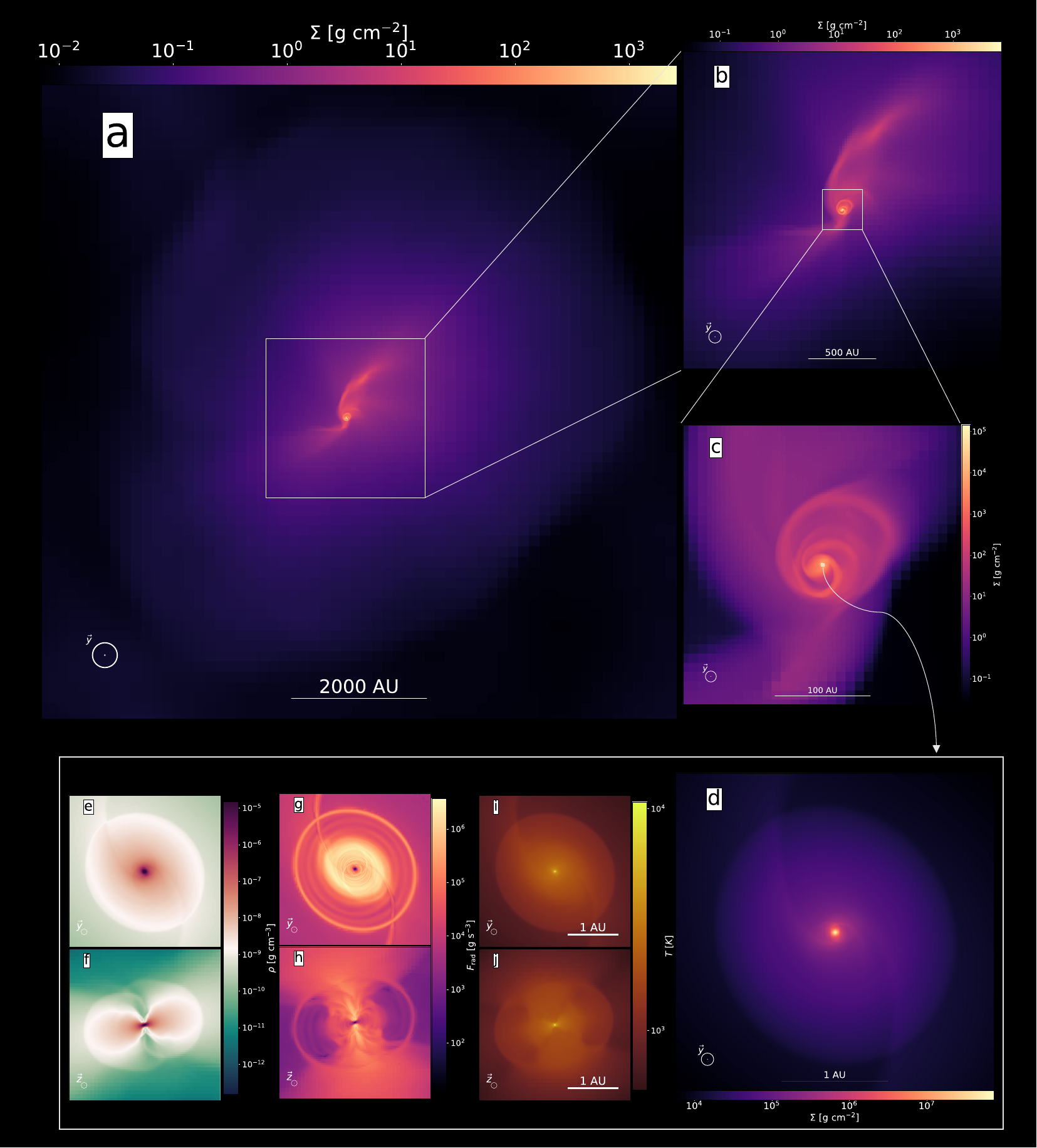}
    \caption{A visualization of the entire dynamical range covered in our simulations. The figure displays data taken from the final snapshot of run G4. Panels (a), (b), (c) and (d) show the projected gas column density along the $\vec{y}$ direction across multiple scales, ranging from $\approx 9.35\times 10^{3}\ \mathrm{AU}$ in panel (a) to $3\ \mathrm{AU}$ in panel (d). Panels (e)-(j) are slices through the center of the domain along the $\vec{y}$ direction for panels (e), (g), and (i), and along the $\vec{z}$ direction for panels (f), (h) and (j). These display density (panels e \& f), radiative flux (panels g \& h), and temperature (panels i \& j). The colorbars in panels (e) \& (f) have been centered on the density of the inner disk's shock front ($\approx 1.5\times 10^{-9}\ \mathrm{g\ cm^{-3}}$ at this snapshot).}
    \label{fig:multiscale}
\end{figure*}

Herein, we illustrate the full dynamical range covered by all of our simulations using run G4 as an example. To this end, we display in \hyperref[fig:multiscale]{Fig. \ref*{fig:multiscale}} column densities at various scales (panels a-d), and slices through the center of the domain displaying density (panels e-f), radiative flux (panels g-h), and temperature (panels i-j). Panel (a) displays the column density at the scale of the dense molecular cloud core. Here, a filamentary structure of size $\simeq 10^{3}\ \mathrm{AU}$ formed by gravo-turbulence can be seen \citep{tsukamoto_2013}, and within this structure a first Larson core is born. In this run, the first core lifetime is significantly extended thanks to ample angular momentum, which reduces the mass accretion rate onto it. As such, a disk was able to form around it\footnote{In the case of runs G1 and G2, no disk was formed at these scales prior to the onset of the second collapse. This is discussed further in \hyperref[section:evol]{Sec. \ref*{section:evol}}.}, as seen in panel (c). Within this disk, the second collapse takes place and gives birth to the protostar and the circumstellar disk, as seen in the lower panels.

\subsection{Protostellar breakup}
\label{section:breakup}
\begin{figure*}
    \centering
    \includegraphics[width=1.\textwidth]{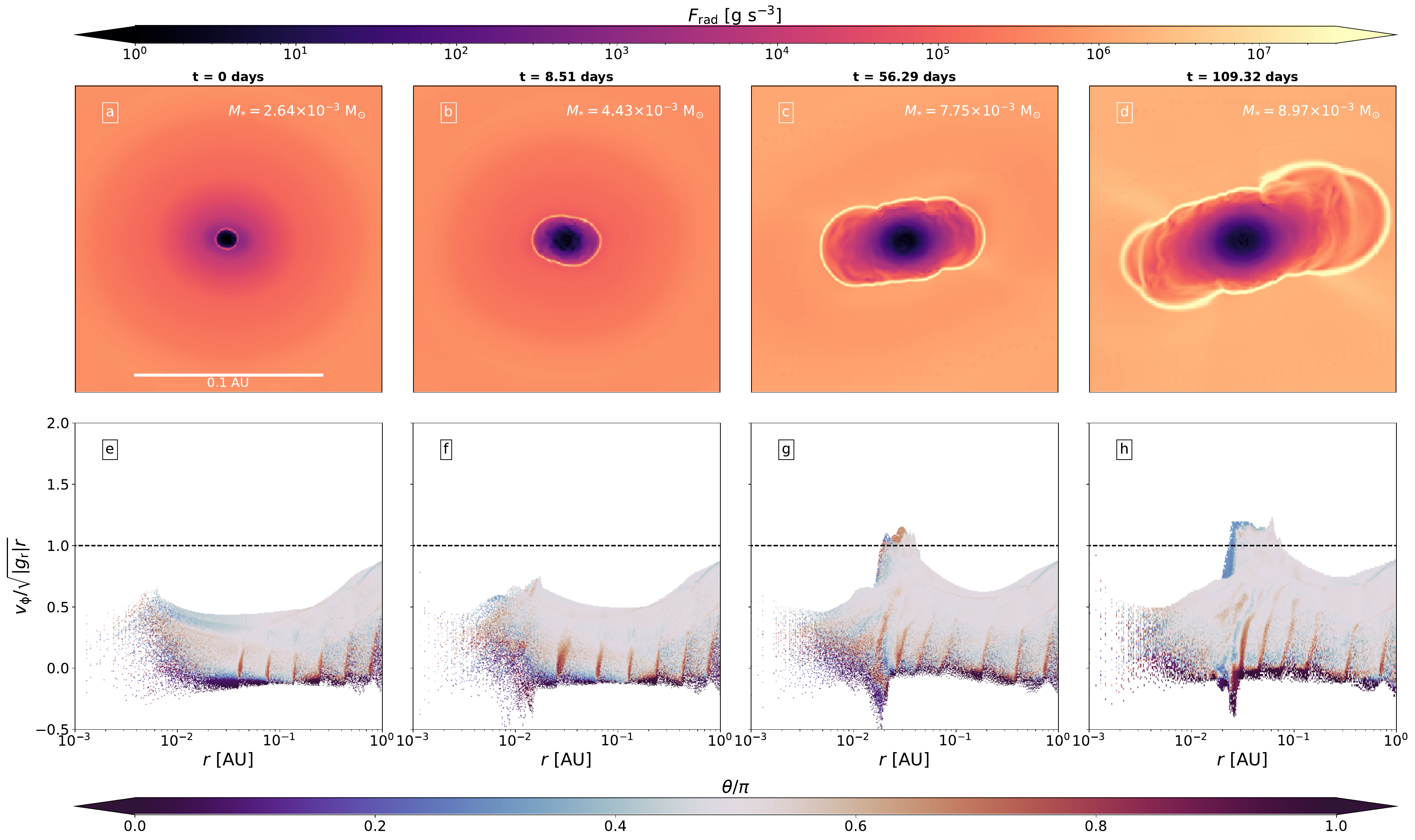}
    \caption{A demonstration of the breakup of the protostar due to excess angular momentum. The data is taken from run G6. Each column represents a different time, where $t=0$ marks the birth of the protostar. Panels (a)-(d) are slices along the $\vec{z}$ direction through the center of the domain which display the local radiative flux emanating from the cells (see \hyperref[eq:Frad]{Eq. \ref*{eq:Frad}}). The scale bar in panel (a) applies to panels (b)-(d) as well. The mass of the protostar at each snapshot is displayed in the top-right corners of panels (a)-(d). Panels (e)-(h) display 2D histograms binning all the cells in our computational domain, which show the distribution of azimuthal velocities divided by $\sqrt{|g_{\mathrm{r}}|r}$ with respect to radius. The color-code in the histograms represents the co-latitude $\theta$ divided by $\pi$, where $\theta/\pi =0.5$ corresponds to the equator, and $\theta/\pi = 1$ (respectively 0) corresponds to the south (respectively north) pole. The dotted black lines in panels (e)-(h) display $v_{\phi}/v_{\mathrm{crit}} =1 $ (see \hyperref[eq:vcrit]{Eq. \ref*{eq:vcrit}}). An animated version of this plot is available in the online journal.}
    \label{fig:breakup}
\end{figure*}

We begin by describing the structure of the system shortly after the onset of the second gravitational collapse using \hyperref[fig:breakup]{Fig. \ref*{fig:breakup}}. We use data from run G6 as an example, although the evolutionary sequence displayed here applies to all other runs as well. The top row of this figure (panels a-d) displays the local radiative flux, an excellent tracer of shock fronts. It is computed as
\begin{equation}
\label{eq:Frad}
F_{\mathrm{rad}} = -\frac{c\lambda\nabla E_{\mathrm{r}}}{\rho\kappa_{\mathrm{R}}},
\end{equation}
where $c$ is the speed of light, $\lambda$ the \cite{minerbo_1978} flux limiter, $E_{\mathrm{r}}$ the radiative energy, $\rho$ the gas density, and $\kappa_{\mathrm{R}}$ the Rosseland mean opacity. At a shock front, kinetic energy is converted into radiation and as such it is accompanied by an increase in radiative flux. Hence, this quantity prominently displays accretion shocks and spiral waves.
\\
The bottom row of the figure (panels e-h) displays the azimuthal velocity\footnote{$v_{\phi}$ was computed along the angular momentum vector of the gas within 0.5 AU.} distribution of all cells in our computational domain with respect to radius, which we have divided by the critical velocity beyond which the centrifugal force exceeds the radial component of the gravitational force:
\begin{equation}
    \label{eq:vcrit}
    v_{\mathrm{crit}} = \sqrt{|g_{\mathrm{r}}|r},
\end{equation}
where 
\begin{equation}
\label{eq:gravr}
    g_{\mathrm{r}}=-\frac{\partial\phi}{\partial r}.
\end{equation}
Here, $\phi$ is the gravitational potential obtained through the Poisson equation. 
\\
Once the gas has completed its dissociation of $\mathrm{H}_{2}$ molecules and ample thermal pressure support is gathered, the second Larson core (i.e., the protostar) is formed. At birth (first column), the protostar is a thermally supported spherical object, and its azimuthal velocities are well below $v_{\mathrm{crit}}$. A mere eight and a half days later (second column), the protostar has nearly doubled in mass, and an equatorial bulge is now visible in panel (b). This is due to the fact that as the protostar accretes, it is also accumulating angular momentum. Nevertheless, it is still rotating below $v_{\mathrm{crit}}$. A month later (third column), the outer shells of the protostar finally exceed $v_{\mathrm{crit}}$, after which material spreads outward and transitions to a differential rotation profile in which the centrifugal force is now the main counterbalance to gravity. The sustained accretion ensures a constant flow of material within the protostar exceeding $v_{\mathrm{crit}}$.

\subsection{An embedded protostar}
\begin{figure*}
    \centering
    \includegraphics[width=1\textwidth]{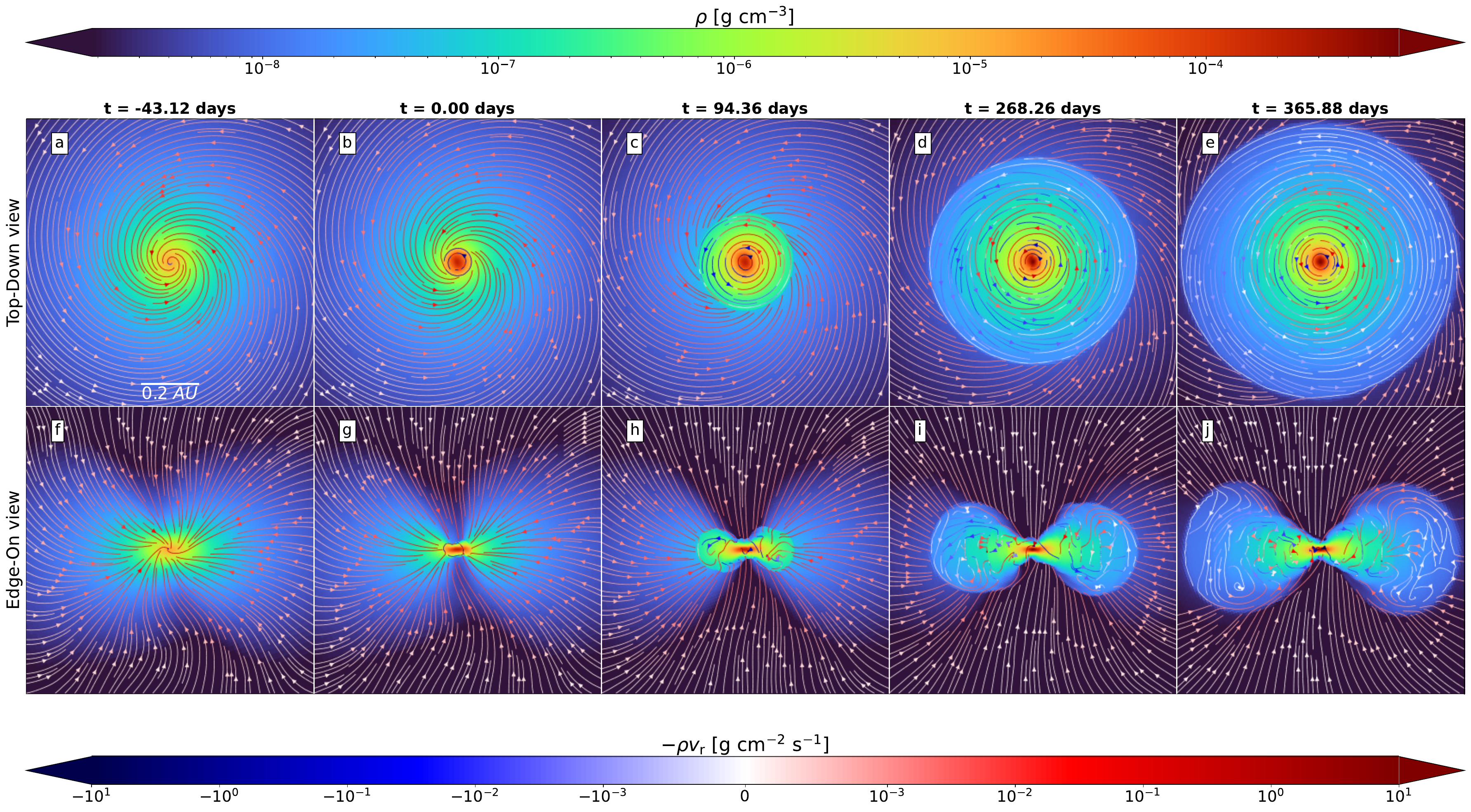}
    \caption{Top-down (top row, panels a-e) and edge-on (bottom row, panels f-j) slices through the center of the domain of run G1, displaying density and velocity streamlines. The color coding in the velocity streamlines displays the local radial mass flux $-\rho v_{\mathrm{r}}$. Each column displays a different epoch, where t=0 (panels b and g) corresponds to the moment of protostellar breakup. The scale bar in panel (a) applies to all other panels. An animated version of this plot is available in the online journal.}
    \label{fig:emStar}
\end{figure*}
\begin{figure*}
    \centering
    \includegraphics[width=1\textwidth]{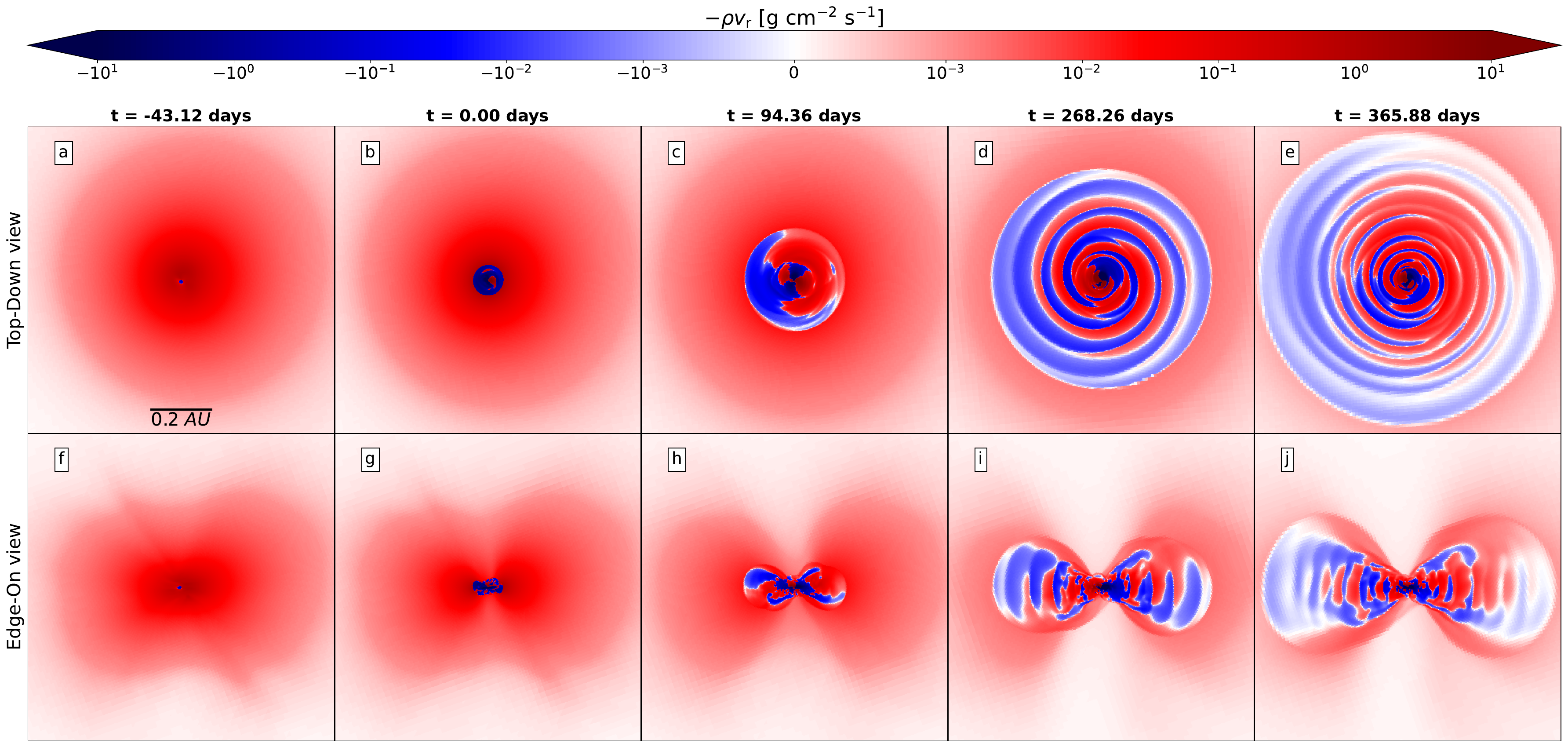}
    \caption{Same as \hyperref[fig:emStar]{Fig. \ref*{fig:emStar}}, but displaying only the local radial mass flux $-\rho v_{\mathrm{r}}$ in order to better demonstrate how accretion occurs in the star and inner disk system. An animated version of this plot is available in the online journal.}
    \label{fig:massflux}
\end{figure*}
\begin{figure}
    \centering
    \includegraphics[width=.35\textwidth]{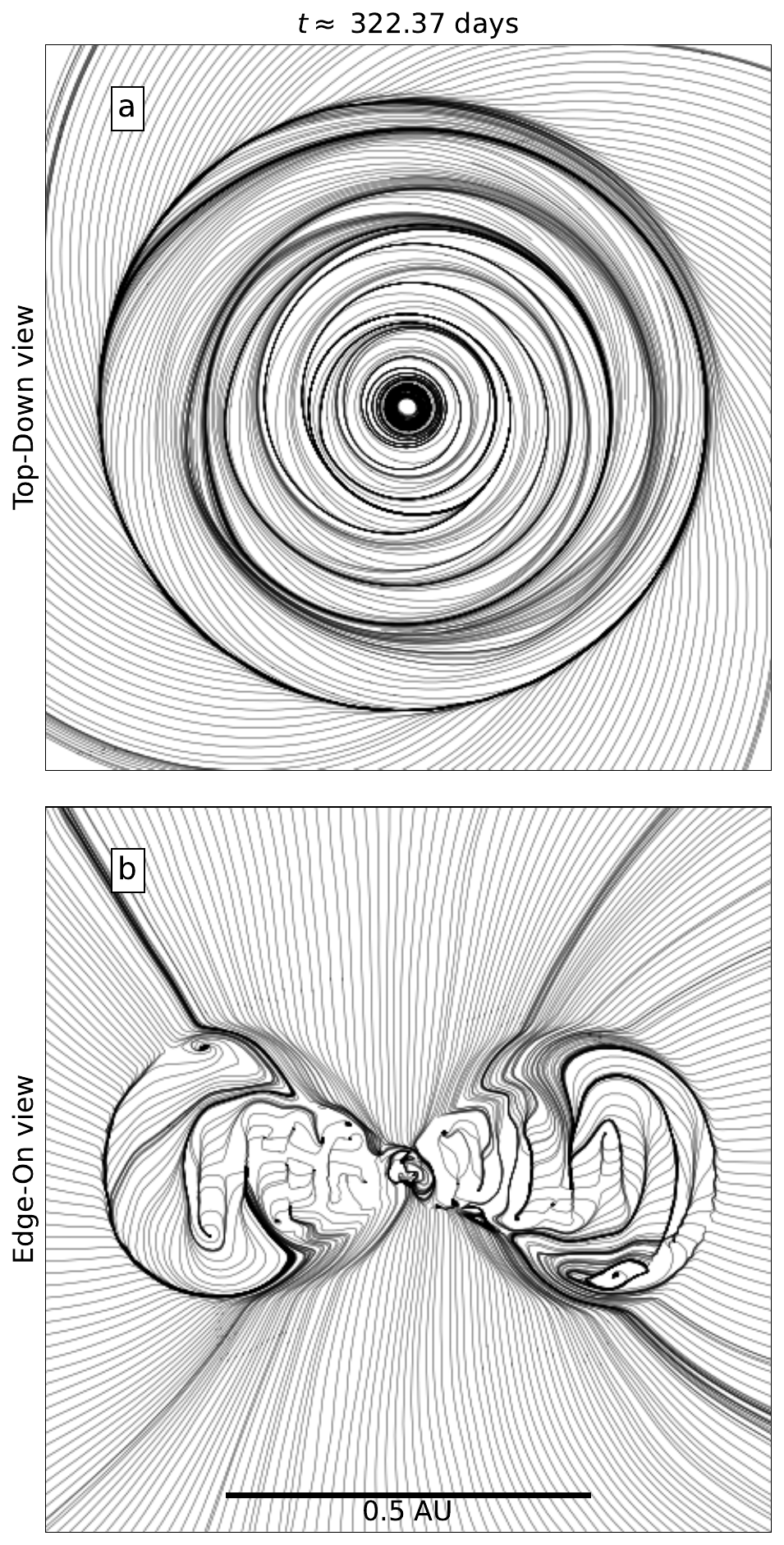}
    \caption{Unbroken velocity vector field streamlines of run G1 at $t\approx 322$ days after the birth of the inner disk, illustrating the dynamics of accretion onto the disk and protostar in a top-down (panel a) and edge-on (panel b) view. The scale bar in panel (b) applies to panel (a) as well. An animated version of these streamlines, made using \href{https://github.com/rougier/windmap}{windmap}, is available in the online journal.}
    \label{fig:wind}
\end{figure}
\begin{figure}
    \centering
    \includegraphics[width=.45\textwidth]{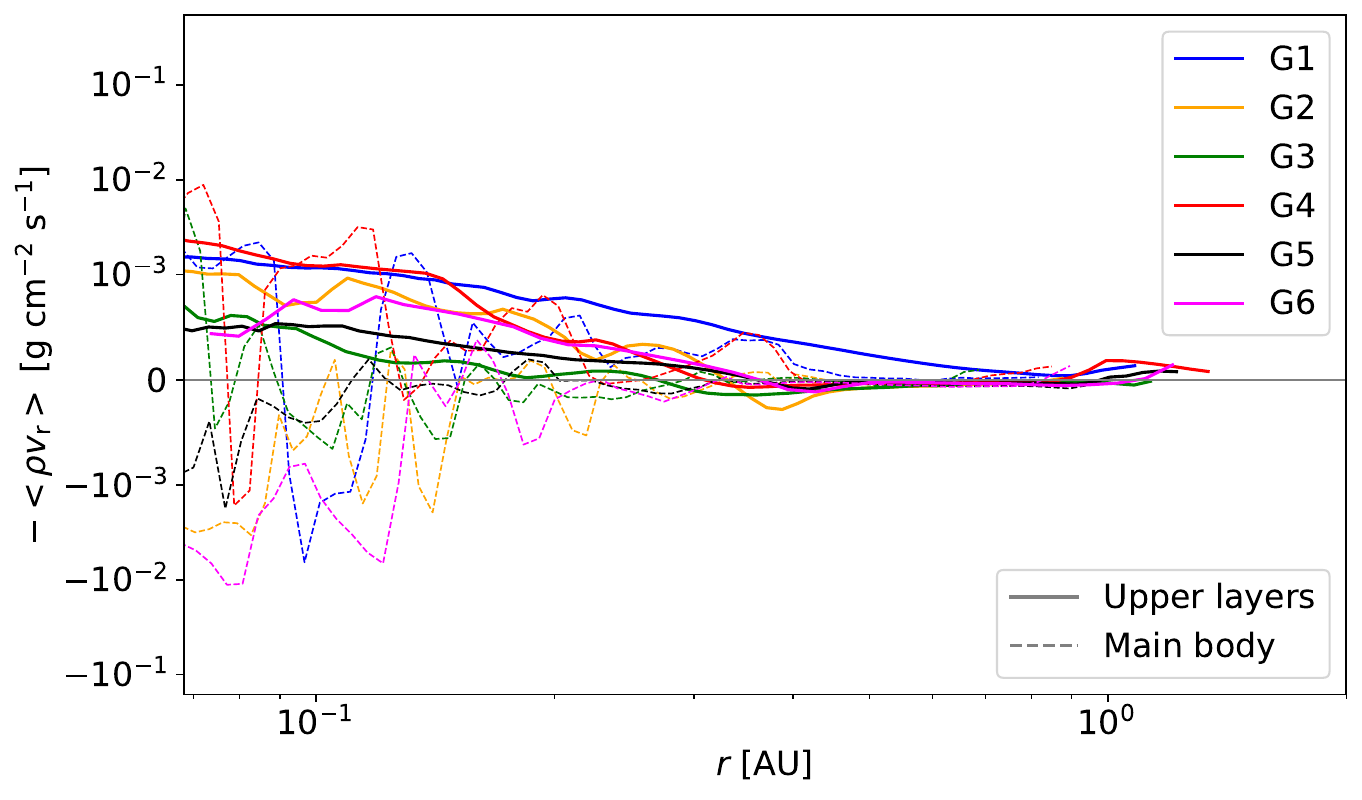}
    \caption{Average radial mass flux measured in radial bins for runs G1 (blue), G2 (orange), G3 (green), G4 (red), G5 (black), and G6 (magenta), at a moment in time where the inner disk has reached $\approx 1\ \mathrm{AU}$ in radius. Only cells belonging to the inner disk were considered (see \hyperref[appendix:definitions]{Appendix \ref*{appendix:definitions}} for information on how the inner disk was defined). The solid lines represent measurements made on the upper layers of the disk, whereas the dashed lines represent measurements made in its main body.}
    \label{fig:dirmassflux}
\end{figure}
\begin{figure}
    \centering
    \includegraphics[width=.4\textwidth]{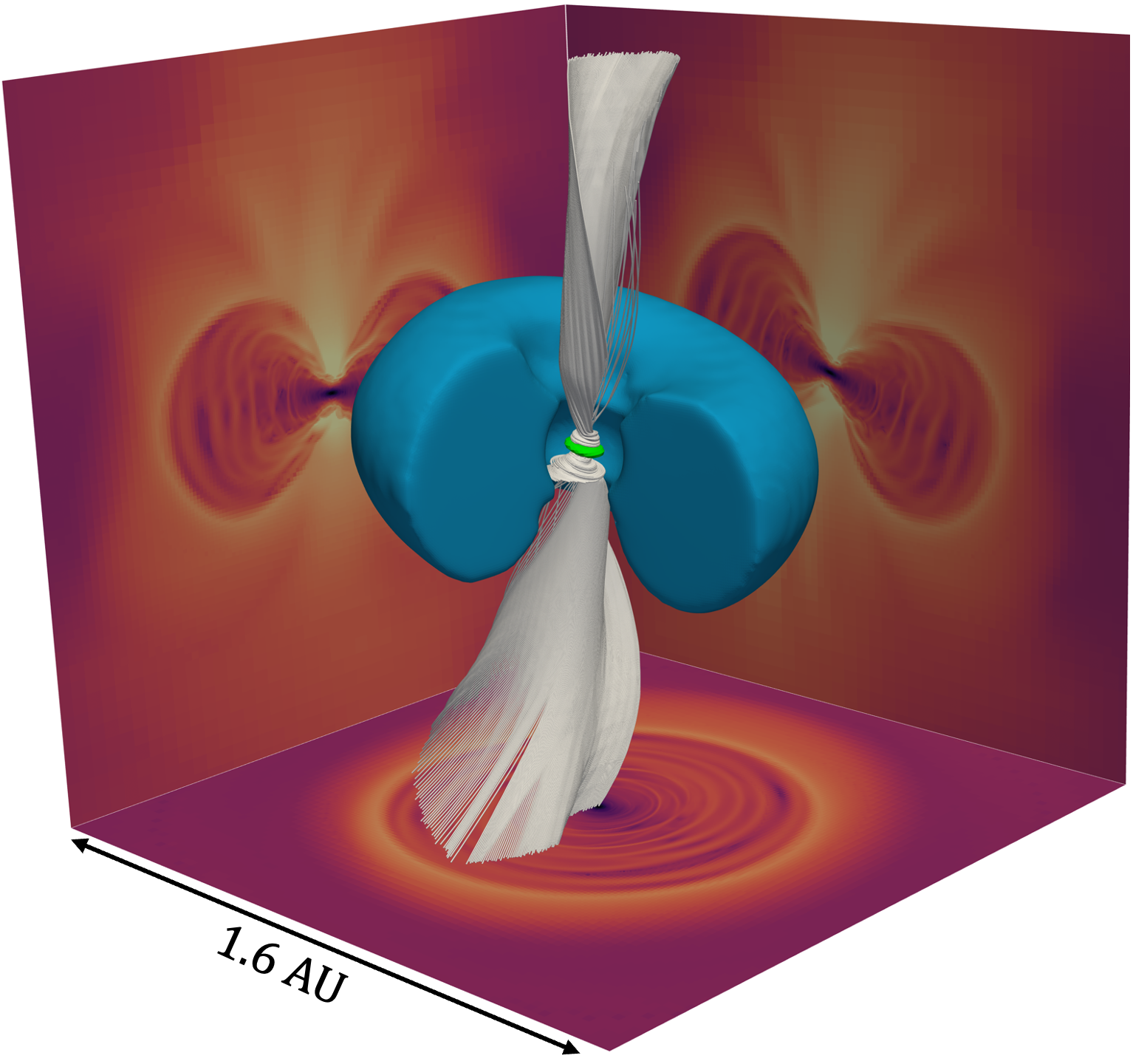}
    \caption{3D view of the inner disk and protostar at a moment in time when the former has reached $\approx 0.5\ \mathrm{AU}$ in radius. The blue structure is the surface of the inner disk. The inner $r<0.1\ \mathrm{AU}$ region has been cut out in order to reveal the flow onto the protostar (rendered in green). The white curves are velocity vector field streamlines, launched along the poles to reveal polar accretion. The bottom, left, and right panels are cross-sections through the center of the domain displaying the radiative flux. The visualized volume is $1.6\times1.6\times1.6\ \mathrm{AU^{3}}$. An animated version of this plot is available in the online journal.}
    \label{fig:threeD}
\end{figure}

The expulsion of material by the protostellar surface will naturally lead to the formation of a circumstellar disk. Here, we will study how such a disk grows and evolves, all the while analyzing the accretion mechanism onto the protostar and the disk. To this end, we study \hyperref[fig:emStar]{Fig. \ref*{fig:emStar}}, which displays density slices through the center of the computational domain for run G1 at different times. The velocity streamlines are color-coded with the local radial mass-flux $-\rho v_{\mathrm{r}}$, where red colors denote inward transport of material, blue colors denote outward transport, and white signifies very little transport. The evolutionary sequence displayed here applies to all other runs. 
\\
Panels (a) and (f) display the system once temperatures exceed 2000 K and the dissociation of $\mathrm{H}_{2}$ is triggered, where the gas spirals inward almost isotropically. The central region accumulates material quickly, and once ample thermal pressure support is gathered, the protostar forms. We display in panels (b) and (g) the structure of the system once the protostar reaches breakup velocity. As the protostar's surface begins expelling material, a disk immediately forms afterward, and as time progresses, an increasing amount of material collides with the disk instead of the protostar, thus causing the former to grow significantly over time. Hereafter, this newly-formed circumstellar disk will be referred to as the inner disk (in accordance with the terminology of \citealp{machida_2011b}). In panel (d), we observe the development of spiral waves, which seem to have subsided into near-circular waves in panel (e)\footnote{The gravitational stability of the inner disk is discussed in \hyperref[section:gravStability]{Sec. \ref*{section:gravStability}}}. Note that during this phase, the accretion timescale of the disk ($M_{\mathrm{d}}/\dot{M_{\mathrm{d}}} \sim 10^{-2}\ \mathrm{M_{\odot}}\ / \ 10^{-2}\ \mathrm{M_{\odot}\ yr^{-1}} = 1\ \mathrm{yr}$) is shorter than its dynamical timescale ($2\pi R_{\mathrm{d}}/v_{\phi}$ $\sim 2\pi\times 1\ \mathrm{AU}\ / \ 3\ \mathrm{km\ s^{-1}} \approx  10\ \mathrm{yr}$). This means that any angular momentum redistribution process within the inner disk occurs on longer timescales than accretion. Thus, accretion is the dominant process behind the expansion of the disk.
\\
We now turn to studying the accretion process with the aid of the streamlines in \hyperref[fig:emStar]{Fig. \ref*{fig:emStar}}, and \hyperref[fig:massflux]{Fig. \ref*{fig:massflux}}, which displays the radial mass flux in slices through the center of the domain. In addition, we display unbroken velocity vector field streamlines in \hyperref[fig:wind]{Fig. \ref*{fig:wind}} at a curated moment. Although the polar regions initially bring a large amount of material to the central protostar, the polar reservoir of gas is quickly depleted and by $t\approx 268\ \mathrm{days}$ (fourth columns and onward of Figs. \hyperref[fig:emStar]{\ref*{fig:emStar}}, \hyperref[fig:massflux]{\ref*{fig:massflux}}), very little mass is accreted through the poles. Indeed, most of the material landing at the protostellar surface is sliding on the inner disk's surface, as its velocity component normal to the disk surface is not strong enough to break through the shock front. We note however that some material landing on the disk surface can sometimes break through the shock front and is then transported into the inner disk, as can be seen in panel (b) of \hyperref[fig:wind]{Fig. \ref*{fig:wind}}. The previously mentioned spiral waves can be seen transporting material radially in panels (d) and (e). This is more apparent in \hyperref[fig:dirmassflux]{Fig. \ref*{fig:dirmassflux}}, which displays the radial mass flux averaged in radial bins and measured separately for both the upper layers of the disk and its main body for all our runs. One can see that in all runs and across all radii, the upper layers of the disk have a strictly positive radial mass flux, whereas the main body shows alternating inflows and outflows of material.
\\
What is most visually striking from the edge-on views is the vertical extent of the disk: it is substantially flared, giving it the shape of a torus. This is more so apparent in the 3D rendering of the system displayed in \hyperref[fig:threeD]{Fig. \ref*{fig:threeD}}. The inner disk's surface (rendered in blue) completely engulfs the protostar (rendered in green). \hyperref[fig:threeD]{Figure \ref*{fig:threeD}} also displays the 3D velocity streamlines, which show that the material accreted through the poles carries with it angular momentum as the gas is spiraling inwards. The cross-sectional slices in the figure display the radiative flux, which reveals shock fronts and spiral waves. Interestingly, there does not seem to be any shock fronts separating the protostar from the inner disk (barring the spiral waves). This means that the accretion shock envelopes both the protostar and the inner disk, and the two act as a continuous fluid system. As such, differentiating the protostar and the inner disk is rather difficult, but the rotational profiles seem to indicate that the protostar is in solid body rotation and the inner disk exhibits differential rotation\footnote{See \hyperref[appendix:definitions]{Appendix \ref*{appendix:definitions}} for an overview of how each object was defined.}. All other runs have displayed an identical structure of the inner disk.

\subsection{A convergent structure}
\label{section:fradvisu}
\begin{figure*}
    \centering
    \includegraphics[width=1.\textwidth]{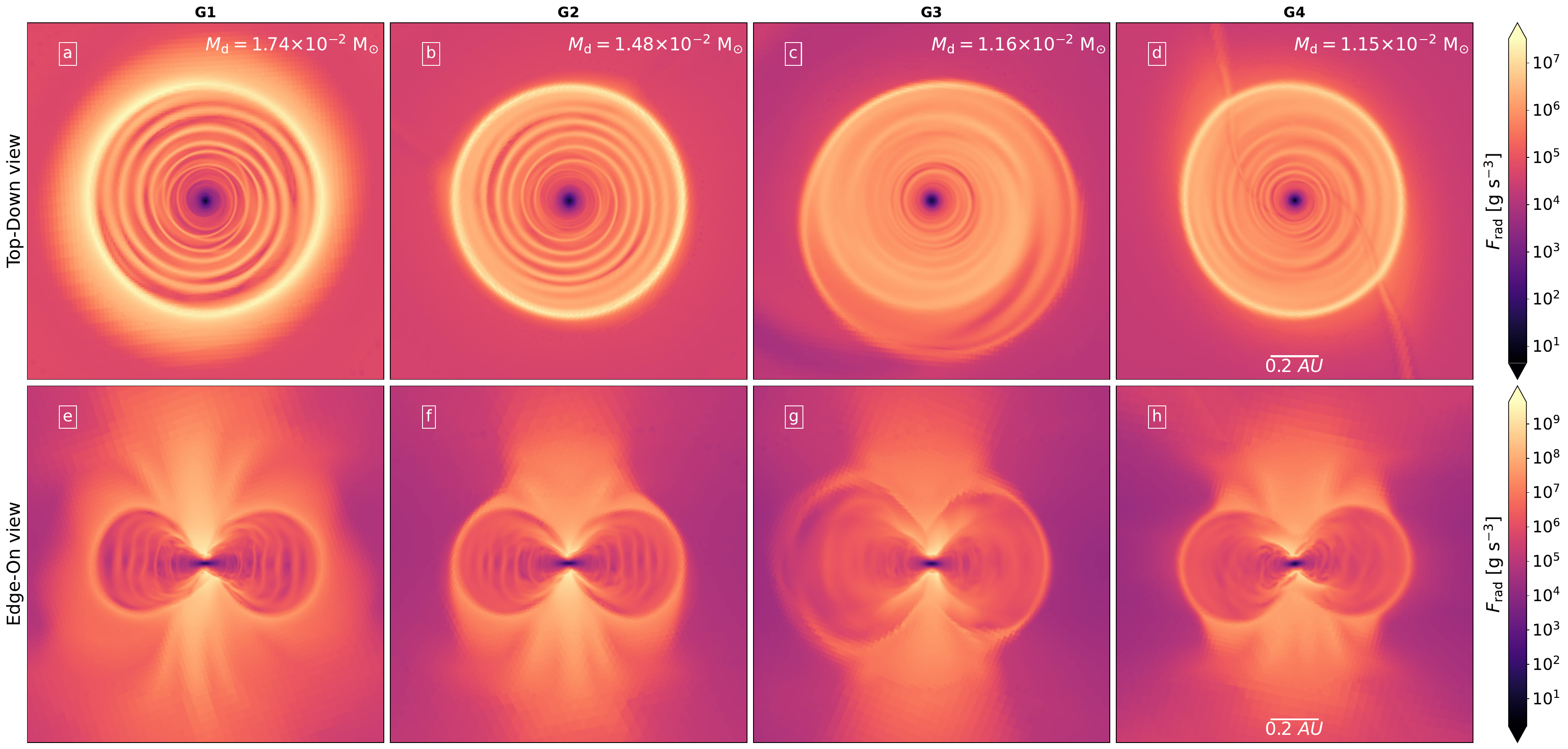}
    \caption{Slices through the center of the domain for runs G1, G2, G3, and G4 (respectively first, second, third, and fourth columns) showing the local radiative flux in a top-down (top row, panels a-d) and edge-on (bottom row, panels e-h) view, which illustrates the structure of the accretion shock as well as the presence of spirals in the inner disk (see \hyperref[eq:Frad]{Eq. \ref*{eq:Frad}}). The scale bar in panels (d) and (h) apply to all other panels. The slices are shown at a moment in time where the inner disks have reached a radius of $\approx$ 0.5 AU. The mass of each inner disk is displayed in the top-right corners of panels (a)-(d).}
    \label{fig:radflux}
\end{figure*}

\begin{figure}
    \centering
    \includegraphics[width=.5\textwidth]{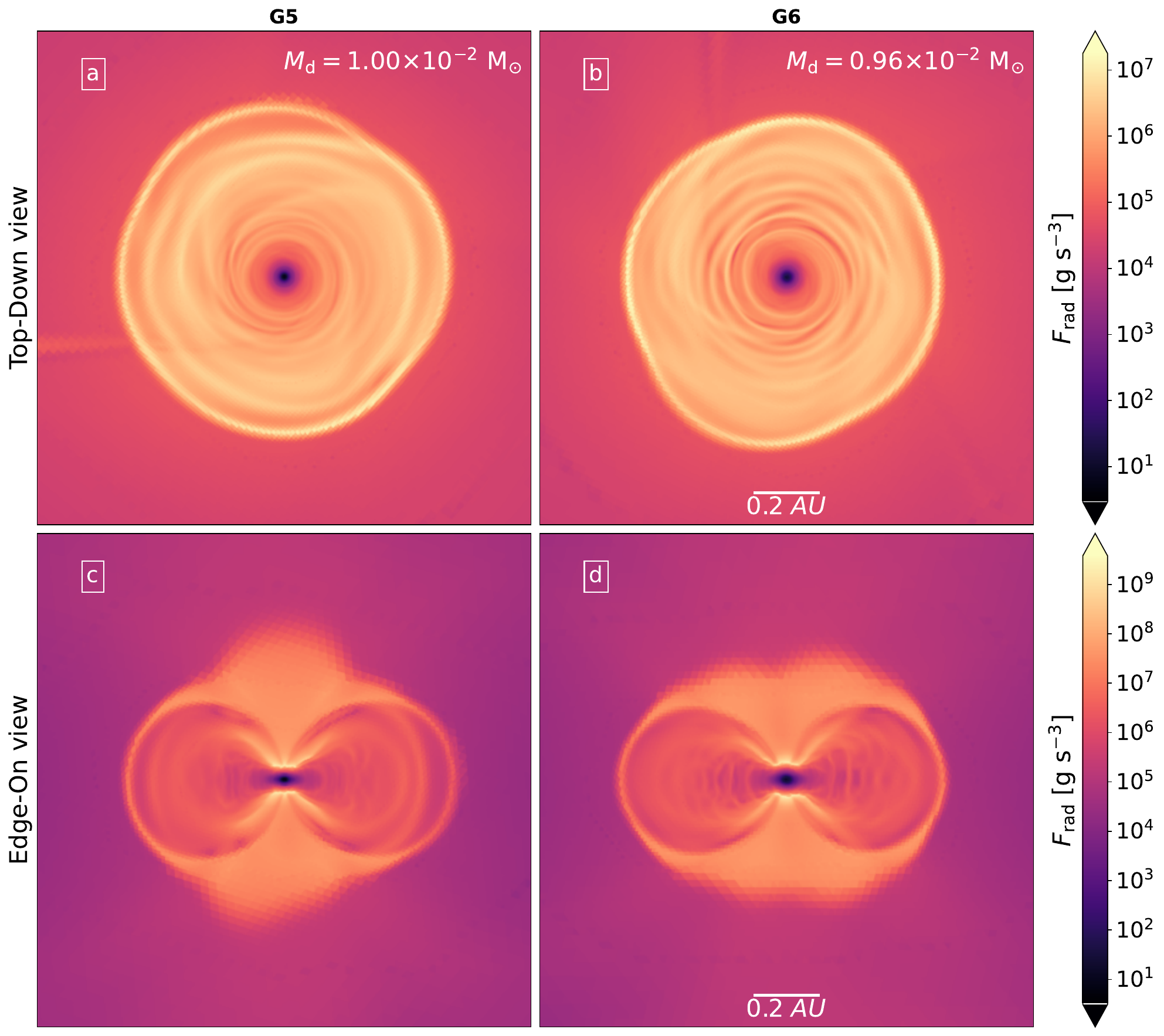}
    \caption{Same as \hyperref[fig:radflux]{Fig. \ref*{fig:radflux}}, but for runs G5 and G6 (respectively first and second columns).}
    \label{fig:radflux2}
\end{figure}

We display in \hyperref[fig:radflux]{Fig. \ref*{fig:radflux}} and \hyperref[fig:radflux2]{Fig. \ref*{fig:radflux2}} the radiative flux in slices through the center of the domain for runs G1-G6 at a moment in time where the inner disk has reached a radius of $\approx 0.5\ \mathrm{AU}$. We note the almost identical structure of the inner disk in all runs: it is toroidal and highly flared. We also note that run G3 (panels c \& q) displays a strong eccentricity, as the outer disk that formed around the first Larson core in this run was already highly eccentric prior to the second collapse.
\\
In the top-down slices (panels a-d), we notice ripples in radiative flux. These are spiral waves which have relaxed into nearly-circular perturbations. We note that runs G1 and G2 have stronger spirals waves (i.e., the radiative flux emanating from them is stronger) due to their higher mass. For a more in-depth analysis of these spiral waves, see \hyperref[section:gravStability]{Sec. \ref*{section:gravStability}}.
\\
An interesting observation from panels (e)-(h) of \hyperref[fig:radflux]{Fig. \ref*{fig:radflux}} is the prominence of the radiative flux along the poles. Indeed, the polar region is much less dense than the equator, which causes the radiative flux to escape much more easily along this direction. This also causes the gas to heat up in the polar direction\footnote{The radiative behavior of the system is studied in \hyperref[section:radbehavior]{Sec. \ref*{section:radbehavior}}.}.

\subsection{Evolution}
\label{section:evol}

\begin{figure*}
    \centering
    \includegraphics[width=1\textwidth]{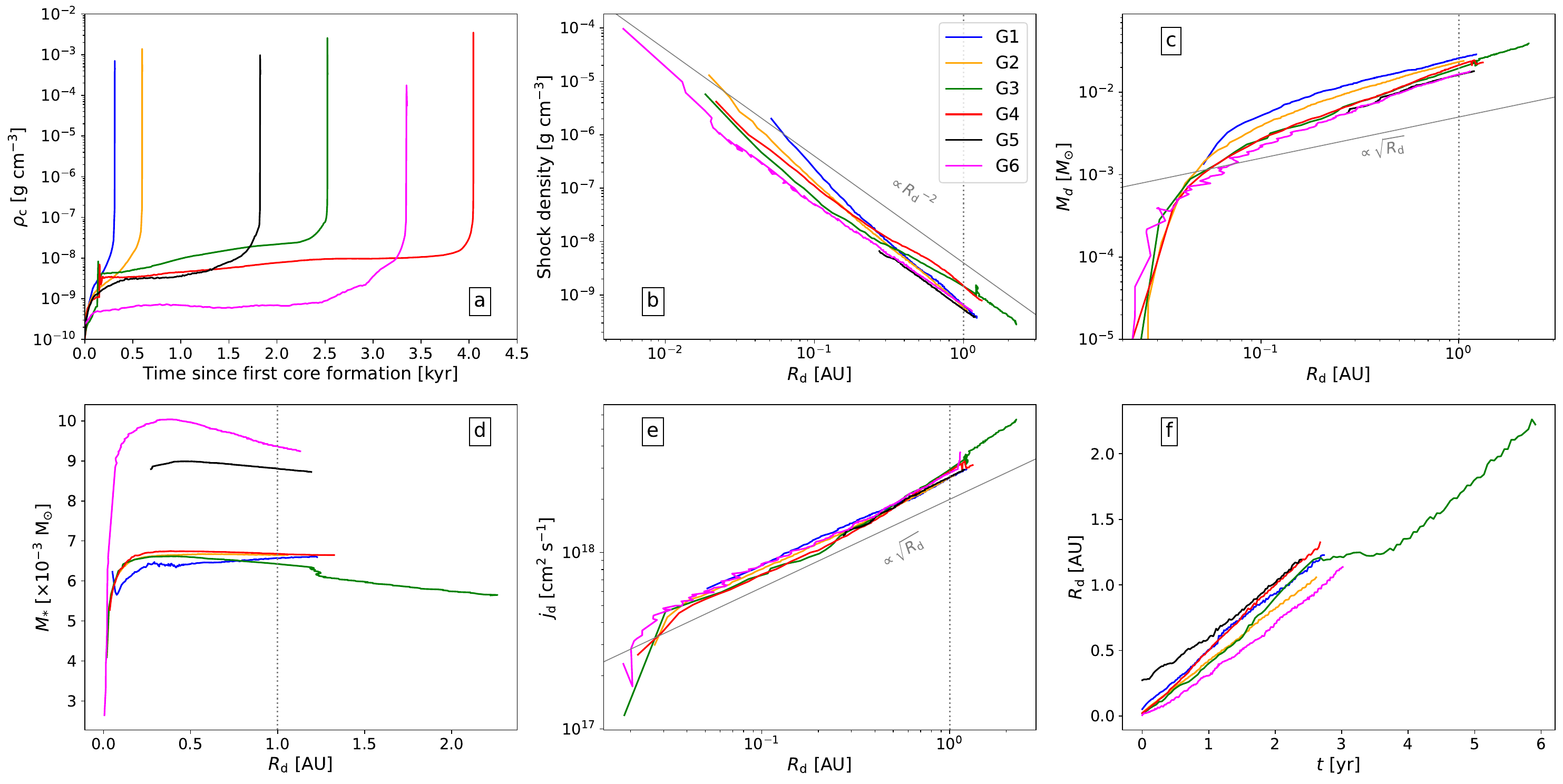}
    \caption{A temporal analysis of the inner disk in all our runs: blue for G1, orange for G2, green for G3, red for G4, black for G5, and magenta for G6. Panel (a) provides the context in which the inner disk is born by displaying the central density's evolution since the formation of the first Larson core (defined as the moment where $\rho_{\mathrm{c}}>10^{-10}\ \mathrm{g\ cm^{-3}}$). Panels (b), (c), (d), and (e) display as a function of the inner disk's equatorial radius $R_{\mathrm{d}}$ (analogous to time), respectively the density of the inner disk's equatorial shock front, the mass of the inner disk, the protostellar mass, and the specific angular momentum of the inner disk. Panel (f) displays the evolution of the inner disk radius with respect to time, where $t=0$ marks the moment of birth of the inner disk. The gray dotted line in panels (b)-(e) marks $R_{\mathrm{d}}=1\ \mathrm{AU}$. For information on how the inner disk was defined, see \hyperref[appendix:definitions]{Appendix \ref*{appendix:definitions}}.}
    \label{fig:diskEvol}
\end{figure*}

Having ascertained the structure of the system in the innermost regions following the second gravitational collapse, we follow the temporal evolution of the inner disk with the aid of \hyperref[fig:diskEvol]{Fig. \ref*{fig:diskEvol}}. The properties of the inner disk once it has reached a radius of $1\ \mathrm{AU}$ are of particular interest, as that is the most commonly chosen sink radius.
\\
First, we point out the different evolutionary history of each simulation in panel (a). This figure displays the maximum density in our computational domain as a function of time since the birth of the first core. The moment in which each curve exhibits a sharp rise in central density corresponds to the onset of the second gravitational collapse. Here, we see clearly that simulations with higher initial amounts of angular momentum  have a delayed onset of second collapse, as the additional centrifugal support significantly extends the first core's lifetime by reducing its mass accretion rate. In runs G3, G4, G5, and G6, the first core lifetime is long enough for it to have a disk built around it, such that the inner disk forms within a disk \citep{machida_2011b}.
Despite the differing evolutionary histories, the resulting properties of the inner disk, displayed in panels (b), (c), (e), and (f), show remarkable similarity. Indeed the temporal evolution of the inner disk equatorial radius $R_{\mathrm{d}}$ (panel f), shows very little spread. The specific angular momentum of the inner disks, displayed in panel (e), also exhibit striking similarity. This shows that the amount of specific angular momentum in the inner disk is independent of the initial amount of angular momentum in the parent cloud core, a result that is in agreement with \cite{wurster_2020}'s non ideal MHD and hydro simulations. Furthermore, the entirety of the angular momentum budget of the first core is found within the inner disk and protostar after it is accreted.
\\
The curves in panel (c) display the inner disk's mass ($M_{\mathrm{d}}$), which exhibit the same evolutionary trend and $M_{\mathrm{d}}(1\ \mathrm{AU})\in [1.634\times 10^{-2}; 2.755\times 10^{-2}]\ \mathrm{M_{\odot}}$. The mass of the protostar (panel d), although very similar in runs G1-4, has runs G5-6 as outliers since $M_{\mathrm{*}}$ is about $\gtrapprox 40\%$ larger in these runs. Interestingly, $M_{\mathrm{*}}$ also seems to be decreasing in most runs, meaning that the protostar is shedding its mass to the disk due to excess angular momentum. The notable exception is run G1, in which the protostar's mass is increasing due to strong gravitational torques.
\\
Finally, we turn our attention to panel (b), which displays the density of the inner disk's equatorial shock front ($\rho_{\mathrm{s}}$). We measure this quantity along the equator since that is the region where most of the incoming mass flux lands on the inner disk (as shown in \hyperref[fig:emStar]{Fig. \ref*{fig:emStar}}). As such, this quantity is an equivalent to the accretion threshold used in sink particles ($n_{\mathrm{acc}}$). We report $\rho_{\mathrm{s}}$($1\ \mathrm{AU})\in [5.35\times 10^{-10}; 2\times 10^{-9}]\ \mathrm{g\ cm^{-3}}$. The most commonly adopted accretion threshold in the literature is $1.66\times 10^{-11}\ \mathrm{g\ cm^{-3}}$ ($n_{\mathrm{acc}} = 10^{13}\ \mathrm{cm^{-3}}$). We thus suggest higher values of $n_{\mathrm{acc}}$ when possible for studies employing sink particles whose radius is $1\ \mathrm{AU}$. However, we acknowledge that this can significantly increase the numerical cost of simulations, and thus might be too constraining for certain studies, particularly those that aim for very long temporal evolution.
\\
\\
The reason for such a convergence in the inner disk properties is the first core itself. As discussed in \cite{ahmad_2023}, this hydrostatic object halts any inward accretion until temperatures can exceed the $\mathrm{H_{2}}$ dissociation temperature of $2000\ \mathrm{K}$, after which the second collapse ensues. As such, the mass accretion rate asymptotically reaches $c_{\mathrm{s}}^{3}/G$ ($\sim 10^{-2}\ \mathrm{M_{\odot}\ yr^{-1}}$, \citealp{larson1969, penston_1969}) independently of initial conditions, provided that a first core forms. The small spread we see in \hyperref[fig:diskEvol]{Fig. \ref*{fig:diskEvol}} is the result of our turbulent initial conditions, as no discernible trend can be inferred from their differences. We expect the large-scale initial conditions to play a more significant role later on when the entirety of the first Larson core is accreted and that mass accretion rates onto the innermost regions are dictated by transport processes within the disk and by the infall of material onto said disk.
\\
\\
We note that in the case where an outer disk exists prior to the onset of a second collapse, the inner disk will simply merge with it \citep{machida_2011b}. This was the case for runs G3, G4 and G5. The first core lifetime in runs G1 and G2 was not long enough for it to gather enough material around it to form a disk.

\subsection{Gravitational stability of the inner disk}
\label{section:gravStability}
\begin{figure*}
    \centering
    \includegraphics[width=1.\textwidth]{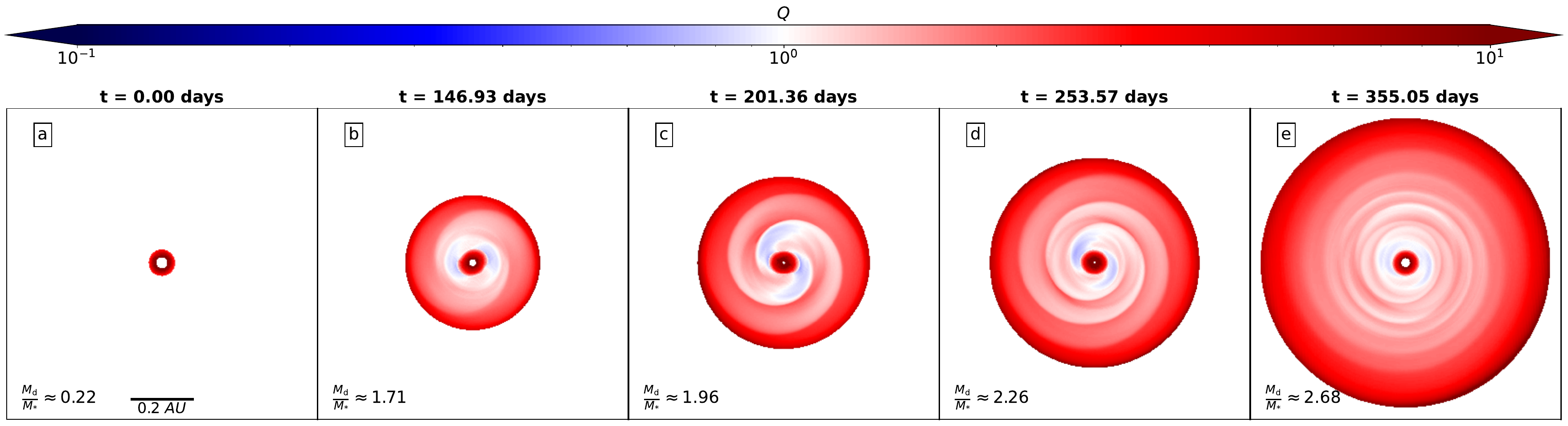}
    \caption{Top-down slices through the center of the domain of run G1, displaying the real part of Toomre $Q$ parameter (see \hyperref[eq:toomre]{Eq. \ref*{eq:toomre}}) at different times, where $t=0$ (panel a) corresponds to the moment of birth of the inner disk. Only cells belonging to the inner disk were used in the calculation of $Q$. The ratio of inner disk mass to protostellar mass is displayed on the bottom left of each panel. The scale bar in panel (a) applies to all other panels. An animated version of this plot is available in the online journal.}
    \label{fig:toomre}
\end{figure*}

The existence of such a massive circumstellar disk naturally begs the question of whether it will undergo fragmentation or not. In order to determine the gravitational stability of the inner disk, we use the classical Toomre $Q$ parameter \citep{toomre_1964}:
\begin{equation}
\label{eq:toomre}
Q = \frac{\omega c_{\mathrm{s}}}{\pi G \Sigma},
\end{equation}
where $c_{\mathrm{s}}$ is the gas sound speed, $\Sigma$ its surface density, and $\omega$ its epicyclic frequency, defined as
\begin{equation}
\omega^{2}=\frac{1}{r^{3}}\frac{\partial (r^{4}\Omega^{2})}{\partial r},
\end{equation}
where $\Omega$ is the angular velocity of the gas. This parameter represents the ratio of the outward pointing forces on the gas, namely the centrifugal and pressure gradient forces, to the inward pointing gravitational force. If $Q<1$, the disk is unstable and a fragmentation is likely. We measure the sound speed and the angular velocity by averaging along the vertical axis of the inner disk.
\\
We display the real part of $Q$ in top-down slices through the center of the domain for run G1 in \hyperref[fig:toomre]{Fig. \ref*{fig:toomre}} at curated moments. In panel (a), we display the circumstellar disk at its birth, just after the protostar exceeded breakup velocity and began shedding its mass. This results in a ring of gas surrounding the protostar, whose $Q$ value is above unity. In panel (b), we display the system just prior to the onset of the first large amplitude spiral wave. Here, the $Q$ parameter remains above unity in the innermost regions of the disk, however an hourglass-shaped region has $Q<1$ at slightly larger radii. The ratio of inner disk mass to protostellar mass has also increased by a factor $\approx 7$. In panel (c), a coherent two-armed spiral wave is launched from the center, and it grows radially as it is sheared apart by differential rotation when propagating outward. Finally, panels (d) and (e) show that these spiral waves relax into nearly circular ripples as a result of the increase in temperature. The $Q$ values in panels (b)-(e) hovers around unity throughout the disk, meaning that the disk remains marginally stable against gravitational instabilities despite its high mass relative to that of the protostar. This is due to its very high temperature.
\\
A recent numerical study by \cite{brucy_2021} has shown that the fragmentation barrier of disks is quite blurry; it is better described by a probabilistic approach and said probabilities strongly depend on how efficiently the disk is able to cool. As such, the fragmentation of the disk is also set by the cooling criterion. In our case, the inner disk is still strongly accreting, and a considerable amount of accretion energy is absorbed by the disk, particularly along the optically thick equator (this will be shown in \hyperref[section:radbehavior]{Sec. \ref*{section:radbehavior}}). This ensures that the inner disk remains hot, and thus, mostly stable against fragmentation despite its very high mass relative to the protostar. The primary regulator of the inner disk temperature during this phase is the endothermic dissociation of $\mathrm{H_{2}}$, which places $\gamma_{\mathrm{eff}}$ at $\approx 1.1$.

\section{Gas structure and kinematics}

Herein, we provide a more quantitative analysis of the structure and kinematics of the inner disk with the aid of \hyperref[fig:oneD]{Fig. \ref*{fig:oneD}}. Panels (a)-(e) of this figure display various quantities azimuthaly averaged in radial bins in the equatorial region, where the equator was defined as the region in which $\theta/\pi \in[0.45;0.55]$, where $\theta$ is the co-latitude. The curves are shown at a moment in time in which the inner disk has reached a radius of $\approx 1\ \mathrm{AU}$.
\\
Panel (a) displays the equatorial density curve, which exhibits a plateau in the innermost regions ($r<10^{-2}\ \mathrm{AU}$) and a power law tail\footnote{We note that the disk in which the protostar was born in run G6 had fragmented prior to second collapse, and 3 first Larson cores exist within it. This causes the density spikes seen at larger radii.}. This behavior has been described by all previous 3D studies in the literature that employ either pure hydro or non-ideal MHD. This density structure, as well as the radial velocity curves displayed in panel (b), suggest that no discontinuity in the flow separates the protostar from the inner disk. The only way for us to differentiate the two is by studying the rotational behavior of the system: we see an object in solid body rotation in the inner-most regions ($j\propto r^{2}$, panel d), and a transition to a differential rotation profile (panel c). 

\subsection{Deviations from Keplerian rotation}

We see a significant deviation from Keplerian rotation ($v_{\mathrm{\phi}}\propto r^{-0.5}$) in the inner disk. Indeed, it seems that $v_{\phi}\propto r^{-0.3}$. In order to explain such difference, one must start by analysing the balance equation between the centrifugal, pressure, and gravitational forces \citep{pringle_1981, lodato_2007}:
\begin{equation}
    \frac{v_{\mathrm{\phi}}^{2}}{r} = \frac{1}{\rho}\frac{\partial P}{\partial r}-g_{\mathrm{r}},
\end{equation}
where $P$ is the thermal pressure. By assuming a radial density profile where $\rho \propto r^{-\beta}$, and radial isothermality ($\partial c_{\mathrm{s}}^{2}/\partial r\approx 0$), we may write
\begin{equation}
\label{eq:vcor}
    v_{\mathrm{\phi}}^{2} \approx -g_{\mathrm{r}}r - \beta c_{\mathrm{s}}^{2},
\end{equation}
where $c_{\mathrm{s}}$ is the gas sound speed. We may approximate $g_{\mathrm{r}}$ from the column density profile of the inner disk:
\begin{equation}
    g_{\mathrm{r}} \simeq -\frac{G M_{\mathrm{enc}}(r)}{r^{2}},
\end{equation}
where $G$ is the gravitational constant and
\begin{equation}
    M_{\mathrm{enc}}(r) = M_{*} + 2\pi\int_{R_{*}}^{r} \Sigma(r') r' dr'.
\end{equation}
Where $\Sigma$ is the disk's surface density. Note that this is merely an analytical estimate of $g_{\mathrm{r}}$, which is different from the true potential computed in \hyperref[eq:gravr]{Eq. \ref*{eq:gravr}}. Now let us assume a column density profile of $\Sigma \propto r^{-\xi}$. This means that the $g_{\mathrm{r}}r$ term in \hyperref[eq:vcor]{Eq. \ref*{eq:vcor}} scales with $r^{-\xi+1}$, whereas the $\beta c_{\mathrm{s}}^{2}$ term scales with $r^{-\beta(\gamma-1)}$ ($T\propto \rho^{\gamma -1}$). From \hyperref[fig:oneD]{Fig. \ref*{fig:oneD}}, we can write $\xi\approx 3/2$ and $\beta\approx 3$. Additionally, $\gamma \approx 1.1$ in the inner disk (panel e of \hyperref[fig:oneD]{Fig. \ref*{fig:oneD}}). Thus, $g_{\mathrm{r}}r\propto r^{-0.5}$ and $\beta c_{\mathrm{s}}^2\propto r^{-0.3}$. This means that we may expect stronger deviations from Keplerian velocity at larger radii in the inner disk. At $R_{\mathrm{d}}=1\ \mathrm{AU}$, we have $M_{\mathrm{d}}\sim 10^{-2}\ \mathrm{M_{\odot}}$ and $c_\mathrm{s}\sim 1\ \mathrm{km\ s^{-1}}$. As such, $\sqrt{1+\beta c_{\mathrm{s}}^{2}/{g_{\mathrm{r}}r}} \approx 0.8$, and we thus expect the deviation from Keplerian rotation to be of the order of $(v_{\mathrm{\phi}}-v_{\mathrm{K}})/v_{\mathrm{K}} \approx -20\ \%$ (where $v_{\mathrm{K}}=\sqrt{GM_{\mathrm{enc}}/r}$). Panel (g) of \hyperref[fig:oneD]{Fig. \ref*{fig:oneD}} shows us that these are reasonable approximations. 
\\
Note that what we call "Keplerian" rotation in the present paper is different from its commonly adopted meaning in the literature. Indeed, the literature defines the Keplerian velocity as $v_{\mathrm{K,lit}}=\sqrt{GM_{*}/r}$, however as the disk's mass exceeds that of the protostar up to seven fold, the measured deviations from $v_{\mathrm{K,lit}}$ would be $\sim 100\%$.

\subsection{Analytical description of the inner disk structure}
\begin{figure*}
    \centering
    \includegraphics[width=1.\textwidth]{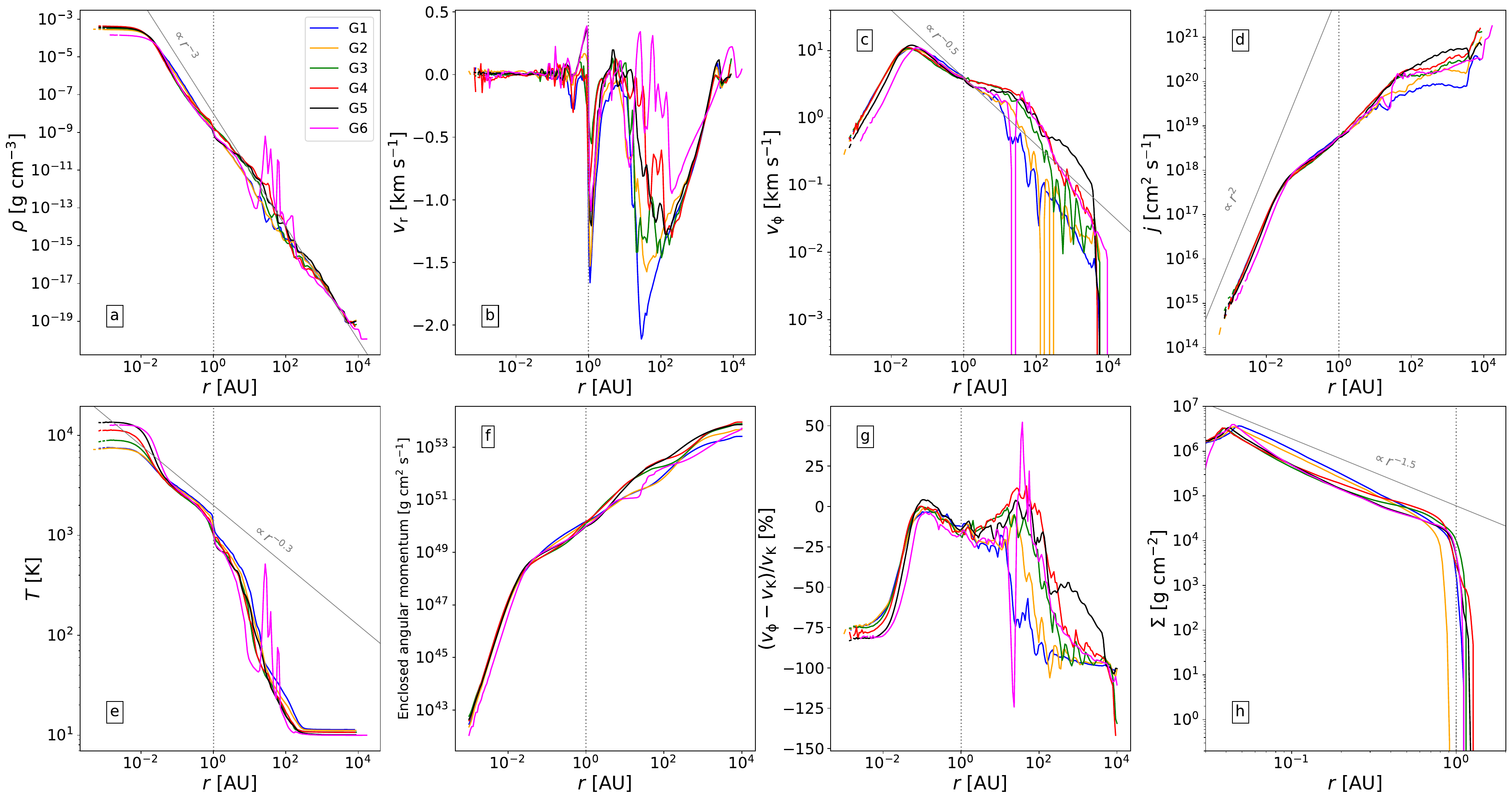}
    \caption{Studying the structure and kinematics of the gas in our simulations. Panels (a)-(e) display a set of azimuthaly averaged quantities along the equatorial regions with respect to radius, where the equator was defined as the region where $\theta/\pi \in [0.45;\ 0.55]$, at a moment in time where the inner disk has reached $\approx 1\ \mathrm{AU}$ in radius. Panel (a) displays density, panel (b) radial velocity, panel (c) azimuthal velocity, panel (d) specific angular momentum, and panel (e) temperature. Panel (f) displays the enclosed angular momentum profile. Panel (g) displays the deviation from Keplerian rotation along the disk midplane. Panel (h) displays the column density profile of the inner disk, where non-disk cells were masked.}
    \label{fig:oneD}
\end{figure*}

Herein, we set out to provide an analytical description of the structure of the inner disk, which is rather exotic by virtue of its very high mass relative to that of the protostar, as well as its very high temperature. This is not meant to be a full analytical development, but rather, one that seeks to deepen our understanding of the structure witnessed in our simulations. Namely, we seek to understand why $\Sigma \propto r^{-3/2}$, and in turn provide an analytical prediction of $n_{\mathrm{acc}}$. To do so, we use as our starting point the important result shown in \hyperref[section:gravStability]{Sec. \ref*{section:gravStability}}, namely that $Q \approx 1$ throughout most of the inner disk. This allows us to link the sound speed $c_{\mathrm{s}}$ and the disk's column density profile $\Sigma$ through
\begin{equation}
\label{eq:link}
c_{\mathrm{s}} \approx \frac{\pi G \Sigma}{\Omega}.
\end{equation}
We can write the following power-law descriptions of these two quantities
\begin{equation}
    \Sigma = \Sigma_0 \left ( \frac{r}{R_{*}}\right )^{-\xi},
\end{equation}
\begin{equation}
    \label{eq:cs}
    c_{\mathrm{s}} = c_{\mathrm{s_{0}}}\left ( \frac{r}{R_{*}} \right )^{-\beta(\gamma - 1)/2},
\end{equation}
where we have made use of the fact that $\rho \propto r^{-\beta}$. Note that $\Omega$ is constrained by $\Sigma$ through the equation
\begin{equation}
\Omega \simeq \sqrt{\frac{G}{r^3}\left ( M_{*}+2\pi\int_{R_*}^{r} \Sigma r' dr' \right )}.
\end{equation}
This allows us to write
\begin{equation}
\Sigma_0 \simeq \frac{c_{\mathrm{s_{0}}}}{\pi G}\sqrt{\frac{GM_*}{R_{*}^3}}.
\end{equation}
To describe $\Sigma$, one must thus first obtain the boundary values at $r=R_{*}$ and $r = 1\ \mathrm{AU}$. Since $\Sigma$ and $c_{\mathrm{s}}$ are linked through \hyperref[eq:link]{Eq. \ref*{eq:link}}, this is equivalent to finding the boundary values of $c_{\mathrm{s}}$. Our simulations indicate $\beta \approx 3$, and $\gamma \approx 1.1$ in the inner disk. Hence, $c_{\mathrm{s}} \propto r^{-0.15}$. This allows us to write
\begin{equation}
\mathrm{log}\left ( \frac{c_{\mathrm{s_{0}}}}{c_{\mathrm{s}}} \right ) =  \frac{\beta(\gamma-1)}{2} \mathrm{log}\left ( \frac{r}{R_*} \right ) = 0.15\times \mathrm{log}\left ( \frac{r}{R_*} \right ).
\end{equation}
Since $R_{*}\sim 10^{-2}\ \mathrm{AU}$,
\begin{equation}
    c_{\mathrm{s_{0}}} = 10^{0.3}\times c_{\mathrm{s}}(1\ \mathrm{AU}) \approx 2\times c_{\mathrm{s}}(1\ \mathrm{AU}).
\end{equation}
This means that when traveling from the protostellar surface to $r=1\ \mathrm{AU}$, $c_{\mathrm{s}}$ reduces by a factor $\approx 2$. The temperature of the inner disk is $\sim 10^{3}\ \mathrm{K}$, hence $c_{\mathrm{s}}(1\ \mathrm{AU}) \sim 1\ \mathrm{km\ s^{-1}}$ and $c_{\mathrm{s_{0}}} \sim 2\ \mathrm{km\ s^{-1}}$. By using \hyperref[eq:link]{Eq. \ref*{eq:link}}, we can link this boundary condition to $\xi$:
\begin{equation}
\left( \frac{c_{\mathrm{s}}}{c_{\mathrm{s_{0}}}} \right)^{2} = \left( \frac{r}{R_{*}} \right)^{3-2\xi} \frac{M_{*}}{M_{\mathrm{enc}}},
\end{equation}
\begin{equation}
\label{eq:ratio}
\Rightarrow \left( \frac{c_{\mathrm{s}}}{c_{\mathrm{s_{0}}}} \right)^{2} = \left( \frac{r}{R_{*}} \right)^{3-2\xi} \frac{M_{*}}{M_{*}+2\pi \frac{\Sigma_{0}}{R_{*}^{-\xi}(-\xi+2)} \left( r^{-\xi+2}-R_{*}^{-\xi+2}\right)}.
\end{equation}
Numerically solving \hyperref[eq:ratio]{Eq. \ref*{eq:ratio}} for $\xi$ such that $(c_{\mathrm{s}}/c_{\mathrm{s_{0}}})^{2}\approx 1/4$ yields $\xi \approx 1.38$ (with $M_{*}\sim 10^{-3}\ \mathrm{M_{\odot}}$), which is close to the value witnessed in most of our runs ($\approx 3/2$, panel h of \hyperref[fig:oneD]{Fig. \ref*{fig:oneD}}).
\\
\\
Now that we have a description of both $\Sigma$ and $c_{\mathrm{s}}$, we may describe the density profile with the aim of predicting $n_{\mathrm{acc}}$ as
\begin{equation}
\rho = \frac{\Sigma}{2h},
\end{equation}
where $h$ is the disk scale height
\begin{equation}
h = \frac{c_{\mathrm{s}}}{\Omega}.
\end{equation}
Thus, this analytical framework provides $\rho(1\ \mathrm{AU}) \approx 5.37\times 10^{-10}\ \mathrm{g\ cm^{-3}}$ ($n_{\mathrm{acc}} \approx 3.23\times 10^{14}\ \mathrm{cm^{-3}}$, when considering a mean molecular weight of 1). This result is within the bounds provided by the simulations ($[5.35\times 10^{-10}; 2\times 10^{-9}]\ \mathrm{g\ cm^{-3}}$), and thus we consider it to be satisfactory.

\subsection{Radial transport within the inner disk}

\begin{figure}
    \centering
    \includegraphics[width=.49\textwidth]{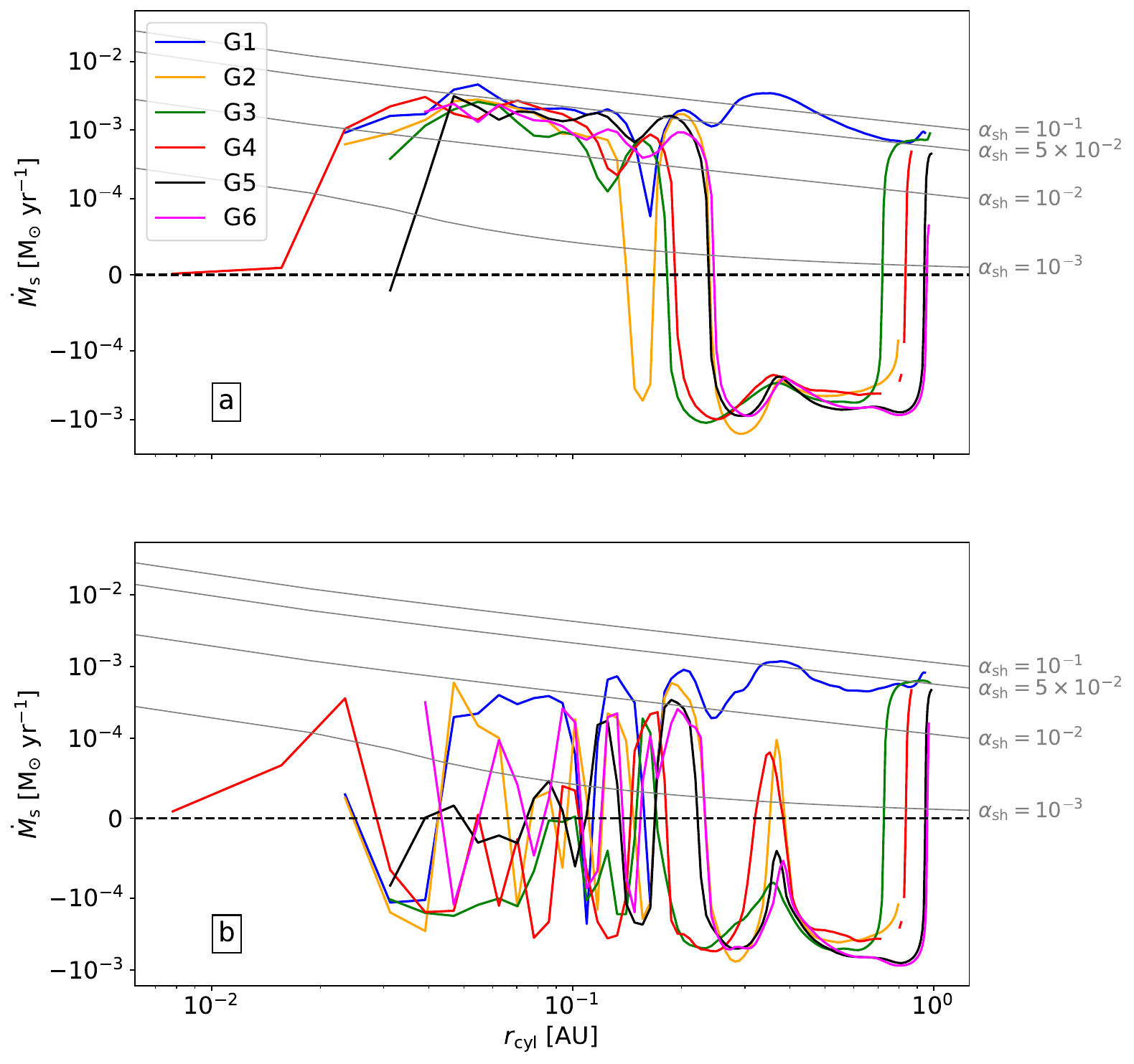}
    \caption{Radial mass accretion rate within the inner disk for all runs (solid colored lines) as a function of cylindrical radius, computed using \hyperref[eq:mdotsim]{Eq. \ref*{eq:mdotsim}} at a moment in time when $R_{\mathrm{d}}$ has reached $\approx 1\ \mathrm{AU}$. Only cells belonging to the inner disk were used in the computation of $\dot{M}_{\mathrm{s}}$. Panel (a) displays the measurements of $\dot{M}_{\mathrm{s}}$ made using ${\langle v_{\mathrm{r}}\rangle}_{z}$ (\hyperref[eq:vrz]{Eq. \ref*{eq:vrz}}), whereas panel (b) displays the same measurements made using ${\langle v_{\mathrm{r}}\rangle}_{\rho}$ (\hyperref[eq:vrrho]{Eq. \ref*{eq:vrrho}}). The solid gray lines are analytical estimates of radial transport, computed using \hyperref[eq:alphamdot2]{Eq. \ref*{eq:alphamdot2}}, with $\alpha_{\mathrm{sh}}=10^{-1}$ (top), $5\times 10^{-2}$ (middle), and $10^{-2}$ (bottom). The other parameters for the solid gray lines are $M_{*}\approx 7\times 10^{-3}\ \mathrm{M_{\odot}}$, $R_{*}\approx 3\times 10^{-2}\ \mathrm{AU}$, $c_{\mathrm{s_{0}}}\approx 5\ \mathrm{km\ s^{-1}}$, $\beta\approx 3$, $\xi\approx 3/2$, and $\gamma\approx 1.1$.}
    \label{fig:alpha}
\end{figure}

Now that we have an analytical framework with which we may describe the inner disk, it is of interest to quantify the transport of material within it. More specifically, we wish to describe transport processes using a simple alpha-disk model \citep{shakura_1973} and compare it with the measured values within our simulations. A common approach in this regard is to measure the stress tensors induced by turbulent fluctuations and the self-gravity of the disk, respectively $\alpha_{\mathrm{R}}$ and $\alpha_{\mathrm{grav}}$ (e.g., \citealp{lodato_2004, brucy_2021, lee_2021}). However, in our case, we have found these measurements difficult to interpret as the disk is awash with eccentric motions (Lovascio et al. 2024 in-prep), turbulent eddies, spiral waves, and erratically infalling material (see for example \hyperref[fig:wind]{Fig. \ref*{fig:wind}}), which resulted in unreliable values of $\alpha_{\mathrm{R}}$ and $\alpha_{\mathrm{grav}}$. Instead, we have opted for a simpler approach in which we attempt to fit an approximate analytical description of the transport within the inner disk with the measured values of our simulations. The mass accretion rate at any given radius in our simulations can be obtained using
\begin{equation}
    \label{eq:mdotsim}
    \dot{M}_{\mathrm{s}}(r) = -2\pi r \Sigma {\langle v_{\mathrm{r}}\rangle},
\end{equation}
where ${\langle v_{\mathrm{r}}\rangle}$ is the vertically averaged radial velocity. Note that the majority of the inward transport of material occurs in the upper layers of the disk, as seen in \hyperref[fig:massflux]{Fig. \ref*{fig:dirmassflux}}. Along the midplane, material tends to spread outward, causing negative mass accretion rates. As such, when performing the vertical average of $v_{\mathrm{r}}$, we weigh $v_{\mathrm{r}}$ in two different ways:
  \begin{equation}
  \label{eq:vrz}
    {\langle v_{\mathrm{r}}\rangle}_{z} = \frac{1}{z_{\mathrm{max}}-z_{\mathrm{min}}} \int_{z_\mathrm{min}}^{z_\mathrm{max}} v_{\mathrm{r}} dz,
  \end{equation}
  \begin{equation}
  \label{eq:vrrho}
    {\langle v_{\mathrm{r}}\rangle}_{\rho} = \frac{1}{\Sigma} \int_{z_\mathrm{min}}^{z_\mathrm{max}} \rho v_{\mathrm{r}} dz,
  \end{equation}
where $z_{\mathrm{min}}$ and $z_{\mathrm{max}}$ are respectively the minimal and maximal heights of the disk at a given radius and azimuth. ${\langle v_{\mathrm{r}}\rangle}_{z}$ will thus be biased by the infall in the upper layers of the disk, whereas ${\langle v_{\mathrm{r}}\rangle}_{\rho}$ will be biased by the dense midplane.

The analytical estimate of the mass accretion rate is \citep{shakura_1973, pringle_1981}
\begin{equation}
    \label{eq:alphamdot1}
    \dot{M}(r) = 3\pi\nu\Sigma,
\end{equation}
where $\nu$ is the effective viscosity of the inner disk
\begin{equation}
    \nu = \frac{\alpha_{\mathrm{sh}}c_{\mathrm{s}}^{2}}{\Omega}.
\end{equation}
Here, $\alpha_{\mathrm{sh}}$ is the \cite{shakura_1973} alpha which dictates the vigour of radial mass transport. Using the equations developed in the previous section, we may write \hyperref[eq:alphamdot1]{Eq. \ref*{eq:alphamdot1}} as
\begin{equation}
    \label{eq:alphamdot2}
    \dot{M}(r) = 3\alpha_{\mathrm{sh}}\frac{c_{\mathrm{s_{0}}}^{3}}{G}\left( \frac{r}{R_{*}} \right)^{-\beta(\gamma-1)-\xi+3/2}\sqrt{\frac{M_{*}}{M_{\mathrm{enc}}}},
\end{equation}
\begin{equation}
\begin{split}
    \Rightarrow \dot{M}(r) = 3 & \alpha_{\mathrm{sh}}\frac{c_{\mathrm{s_{0}}}^{3}}{G}\left( \frac{r}{R_{*}} \right)^{-\beta(\gamma-1)-\xi+3/2}\\
    &\times\sqrt{\frac{M_{*}}{M_{*}+2\pi \frac{\Sigma_{0}}{R_{*}^{-\xi}(-\xi+2)} \left( r^{-\xi+2}-R_{*}^{-\xi+2}\right)}}.
\end{split}
\end{equation}
We may obtain an estimate of the $\alpha_{\mathrm{sh}}$ parameter in \hyperref[eq:alphamdot2]{Eq. \ref*{eq:alphamdot2}} with the aid of \hyperref[fig:alpha]{Fig. \ref*{fig:alpha}}, which displays the mass accretion rate in our simulations computed using \hyperref[eq:mdotsim]{Eq. \ref*{eq:mdotsim}} at a moment in time when the inner disk radius has reached $\approx 1\ \mathrm{AU}$. We measure $\dot{M}_{\mathrm{s}}$ using both ${\langle v_{\mathrm{r}}\rangle}_{z}$ (panel a) and ${\langle v_{\mathrm{r}}\rangle}_{\rho}$ (panel b).
Using average stellar parameters reported in the figure caption, we over-plot \hyperref[eq:alphamdot2]{Eq. \ref*{eq:alphamdot2}} (solid gray lines). The figure shows that the vigour of inward transport varies across all radii, and in the case of runs G2-6, the outer layers ($r>0.2\ \mathrm{AU}$) of the inner disk have negative mass accretion rates, meaning that material is mostly spreading outwards. In panel (b), we see that the transport of material in the main body of the inner disk fluctuates wildly. Furthermore, any inward transport in this region consistently has lower mass accretion rates than in the upper layers of the disk (panel a). Indeed, the upper layers of the disk have a mass accretion rate that can be approximated by \hyperref[eq:alphamdot2]{Eq. \ref*{eq:alphamdot2}} with $\alpha_{\mathrm{sh}}\sim 5\times 10^{-2}$.
\\
Note however that $\dot{M}_{\mathrm{s}}$ varies not only in space but also in time, and so the value of $\alpha_{\mathrm{sh}}$ reported here is not valid throughout the entirety of the class 0 phase. Indeed, once most of the remnants of the first Larson core are accreted, and that the mass accretion rate onto the star-disk system significantly reduces, this will inevitably cause a decrease in $\alpha_{\mathrm{sh}}$, as $\dot{M}_{\mathrm{s}}$ is mainly dominated by the infall and will thus drop by several orders of magnitude. We would thus have a disk whose angular momentum transport is very weak, with $\alpha_{\mathrm{sh}}\sim [10^{-5};10^{-3}]$, but whose mass is significantly higher than that of the protostar. In addition, panel (d) of \hyperref[fig:diskEvol]{Fig. \ref*{fig:diskEvol}} shows that despite the inward transport of material within the inner disk, the protostellar mass is decreasing in most runs as a result of excess angular momentum. This allows us to conclude that although the inner disk can be described by the physics of alpha disks, such a description is a first-order approximation whose results are not entirely reliable.

\section{Radiative behavior}
\label{section:radbehavior}

\begin{figure*}
    \centering
    \includegraphics[width=1.\textwidth]{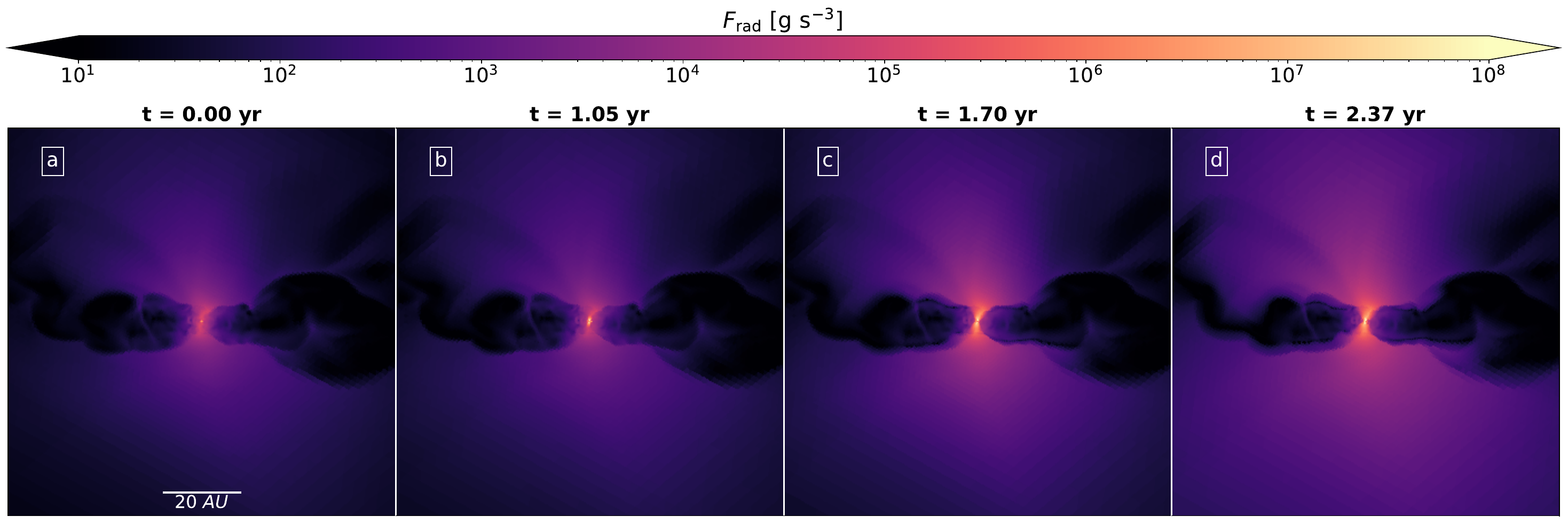}
    \caption{Edge-on slices through the center of the domain displaying the local radiative flux at different times (see \hyperref[eq:Frad]{Eq. \ref*{eq:Frad}}), where $t=0$ (panel a) corresponds to the moment of protostellar birth. The data is taken from run G5. The scale bar in panel (a) applies to all other panels. An animated version of this plot is available in the online journal.}
    \label{fig:fradlargescale}
\end{figure*}

\begin{figure*}
    \centering
    \includegraphics[width=.8\textwidth]{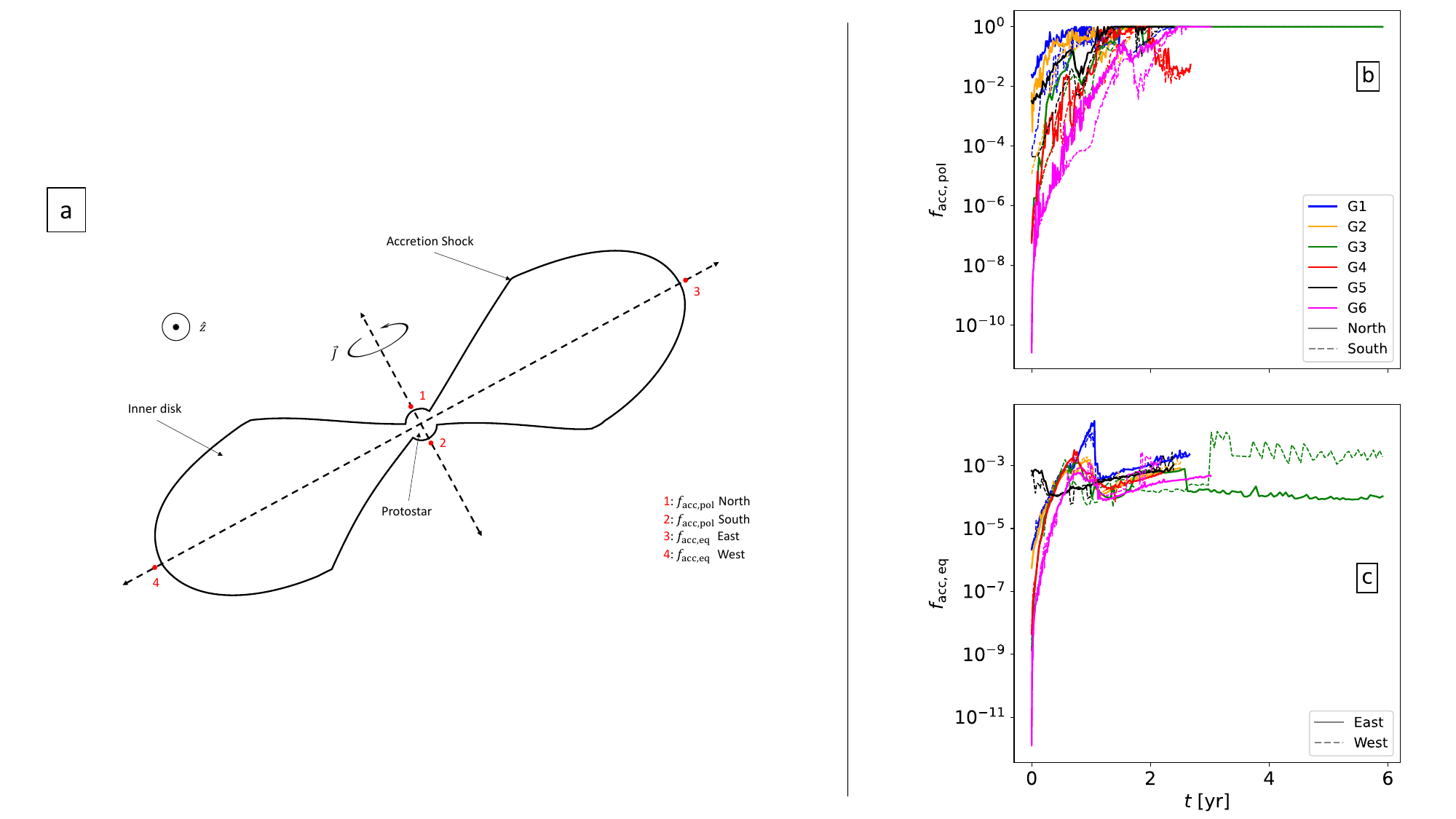}
    \caption{A quantitative analysis of the radiative behavior of the accretion shock. Panel (a) is a schematic representation of the star and inner disk system, where the solid line is the location of the accretion shock that envelopes both the protostar and the disk. The dashed lines represents four rays launched from the center of the system, on which the location of the accretion shock is found and its radiative efficiency measured just upstream from it (numbered red dots). The resulting measurements are presented in panels (b) and (c): north and south (respectively solid and dashed lines in panel b), and east and west (respectively solid and dashed lines in panel c) for all simulations. These display the polar and equatorial radiative efficiencies (respectively $f_{\mathrm{acc,pol}}$ and $f_{\mathrm{acc,eq}}$, see \hyperref[eq:faccpol]{Eq. \ref*{eq:faccpol}} and \hyperref[eq:facceq]{Eq. \ref*{eq:facceq}}).}
    \label{fig:facc}
\end{figure*}

Although the protostar has not reached the temperatures required for fusion yet ($>10^{6}\ \mathrm{K}$), it will dominate the radiative output of the system by virtue of its high temperature and its accretion luminosity. A quantitative analysis of said radiative behavior is seldom provided in the literature, and it is the purpose of this section.
\\
\\
To this end, we first begin by providing a qualitative overview of the radiative behavior of the system at medium (i.e., $10^{2}\ \mathrm{AU}$) scales. As discussed previously, the structure of the accretion shock is rather complex, as it envelopes both the protostar and the inner disk. Nevertheless, the polar accretion shock, as seen in the bottom rows of Fig. \ref{fig:radflux} and \ref{fig:radflux2}, produces the majority of the radiative flux. In addition, the density cavity along the poles allows said radiation to escape much more easily than along the optically thick equator. This is reflected in \hyperref[fig:fradlargescale]{Fig. \ref*{fig:fradlargescale}}, which displays the radiative flux emanating from each cell in edge-on slices across the center of our domain at curated moments for run G5. The protostar is born embedded within a disk that formed around the first Larson core (panel a). As time progresses, the radiation produced at the protostellar accretion shock escapes and brightens the polar regions significantly (panels b, c, and d). However, the equatorial regions remain dark as the radiative flux struggles to pierce through the highly opaque disk. As such, the disk is almost unaffected by the protostar's radiation.
\\
\\
We now turn to providing a quantitative analysis of the radiative behavior of the protostar and inner disk. To do so, as in \cite{vaytet_2018} - \cite{bhandare_2020} - \cite{ahmad_2023}, we compare two quantities: the radiative flux just upstream of the accretion shock ($F_{\mathrm{rad}}$, see \hyperref[eq:Frad]{Eq. \ref*{eq:Frad}}), and the incoming accretion energy flux. However, since the shock front's structure is complex, we measure these quantities in only four directions: north-south along the poles, and east-west along the equator (see panel a of \hyperref[fig:facc]{Fig. \ref*{fig:facc}}):
\begin{equation}
F_{\mathrm{acc},pol} \simeq -\rho v_{\mathrm{r}}\frac{G M_{\mathrm{enc}}(R_{*})}{R_{*}} - E v_{\mathrm{r}} - P v_{\mathrm{r}} - \frac{\rho v_{\mathrm{r}}^{3}}{2},
\end{equation}
\begin{equation}
F_{\mathrm{acc},eq} \simeq -\rho v_{\mathrm{r}}\frac{G M_{\mathrm{enc}}(R_{\mathrm{d}})}{R_{\mathrm{d}}} - E v_{\mathrm{r}} - P v_{\mathrm{r}} - \frac{\rho v_{\mathrm{r}}^{3}}{2},
\end{equation}
where $R_{*}$ (resp. $R_{\mathrm{d}}$) is the protostellar (resp. inner disk) radius, $E$ the gas internal energy, $P$ its thermal pressure, and
\begin{equation}
 M_{\mathrm{enc}}(r) = 4\pi \int_{0}^{r} \rho {r'}^{2}dr'.
\end{equation}
This allows us to define the radiative efficiency as
\begin{equation}
\label{eq:faccpol}
f_{\mathrm{acc,pol}} \approx \frac{F_{\mathrm{rad,pol}}}{F_{\mathrm{acc,pol}}},
\end{equation}
\begin{equation}
\label{eq:facceq}
f_{\mathrm{acc,eq}} \approx \frac{F_{\mathrm{rad,eq}}}{F_{\mathrm{acc,eq}}}.
\end{equation}
These are approximate measurements of the radiative efficiency of the accretion shock because the radiative flux also contains the cooling flux emanating from the protostellar interior, although the later remains smaller than that produced at the accretion shock because of the low temperature of the protostar ($\sim 10^{4}\ \mathrm{K}$) prior to deuterium burning.
\\
We display in panels (b) and (c) of \hyperref[fig:facc]{Fig. \ref*{fig:facc}} the resulting measurements of $f_{\mathrm{acc,pol}}$ and $f_{\mathrm{acc,eq}}$ obtained through ray-tracing. A schematic drawing displaying the location at which each quantity was measured is presented in panel (a). In contrast to spherically symmetrical calculations, the radiative efficiency of the accretion shock displays a strong anisotropy: the polar accretion shock (panel b) is much more efficient than its equatorial counterpart (panel c). Indeed, the polar shock front reaches supercriticality ($f_{\mathrm{acc,pol}}=1$) in less than 2.5 years for most runs, whereas $f_{\mathrm{acc,eq}}<1$ throughout all simulations. This is due to the polar density cavity that allows the accretion shock to shine into an optically thin medium, whereas the equatorial accretion shock remains optically thick due to the remnants of the first Larson core and the presence of an extended outer disk around it.
\\
The drop in $f_{\mathrm{acc,eq}}$ seen in most runs (e.g. at $t\approx 1.2$ yr for run G1) is due to the equatorial shock front expanding out of the opacity gap, a region where temperatures are high enough to sublimate dust (see Figure 1 of \citealp{ahmad_2023}). This causes the shock front to shine into a region of higher opacity, which in turn reduces its luminosity. Eccentric inner disks can also cause an east-west anisotropy, as is most apparent in run G3. In contrast, we witness very minor anisotropies when comparing north-south radiative efficiencies.
\\
The low radiative efficiency of the equatorial shock front has an important consequence on the structure of the inner disk: as it accretes, the majority of the accretion energy is dumped into its thermal budget, thus causing it to maintain a high temperature and to swell along the vertical extent.


\section{Discussions}

\subsection{The star-disk structure}
In the common star formation paradigm, the protostar is often seen as a standalone object wholly separate from the circumstellar disk due to an accretion shock separating the two. Our results show that such a boundary does not exist: the transition from the protostar to the inner disk is a smooth one, and the two act as a continuous fluid system. This means that until either torque mechanisms transport sufficient material inward, or a protostellar magnetosphere arises, the protostar and the inner disk will continue to behave as a continuous fluid system.
\\
Unfortunately, obtaining observational constraints on the structure of the system in such a young and deeply embedded object is very difficult, but two recent studies by \cite{laos_2021} and \cite{gouellec_2024} seem to be offering hints on the accretion mechanism during the class 0 phase. Whereas \cite{laos_2021} argues for magnetospheric accretion similarly to class I objects by basing their arguments on the similar shapes of the Br${\gamma}$ line profiles, \cite{gouellec_2024} additionally analysed the velocity profiles of these lines which exhibited clear differences. Based on this result, they concluded that the accretion mechanism in class 0 sources must be of a different nature. Additionally, they report very vigorous accretion unto the central regions. These observations are an indirect probe of the structure of the system following the second collapse. If a stellar magnetosphere cannot be inferred from them, then we argue that a structure similar to that described in this paper may be present.

\subsection{Toward an eventual fragmentation}
Although our analysis of the gravitational stability of the inner disk indicates that it is currently marginally stable against gravitational instabilities, it is also in a transient state in which it maintains an incredibly high temperature through accretion. This is by virtue of the low radiative efficiency of the inner disk's shock front. The protostar, as shown in \hyperref[section:radbehavior]{Sec. \ref*{section:radbehavior}}, contributes almost nothing to heating the inner disk because it is currently embedded within it and the majority of its radiative flux escapes along the poles. One can imagine that such a disk will inevitably cool down once accretion subsides, and thus would be much more prone to fragmentation. The high mass budget of the inner disk also indicates that such fragmentations can occur multiple times, and thus lead to the formation of multiple star systems, or perhaps of Jupiter-like planets in close proximity to the central star.

\subsection{Caveats}
Although the radiative-hydro approximation is valid for the timescales described here, this obviously cannot be the case on longer timescales, as the stellar magnetic field will undoubtedly increase once a dynamo process begins. Furthermore, the hot, dense, and highly turbulent inner disk may also generate a dynamo process \citep{balbus_1991, wardle_1999, lesur_2014, riols_2019, deng_2020}. The onset of strong magnetic fields will bring about outflows and jets, as well as induce strong magnetic torques that can transport material inward and also reduce the rotation rate of the protostar to observed values in more evolved systems ($\sim 10\%$ of breakup velocity, \citealp{Hartmann_1989, herbst_2007}). These will likely heavily influence the evolution of the system in the innermost regions. Nevertheless, the simulation presented in \cite{machida_2019} was integrated for 2000 years after protostellar birth with non-ideal MHD, and their results seem to indicate that $\rho_{\mathrm{s}}(1\ \mathrm{AU})\sim 10^{-9}\ \mathrm{g\ cm^{-3}}$ (see their Fig. 13). Their protostar also seems to progressively separate from the inner disk (see panels (m)-(p) of their Fig. 5 and $v_{\mathrm{r}}$ in their Fig. 11), likely as a result of magnetic torques. Although state of the art non-ideal MHD simulations indicate that thermal pressure support far outweighs its magnetic   counterpart in circumstellar disks (e.g., \citealp{machida_2010, vaytet_2018, machida_2019, lee_2021}), magnetic torques likely outweigh turbulent viscosity or gravitational torques (as quantified by \citealp{machida_2019}).
\\
In addition, although we notice relatively little spread in our results, more exhaustive constraints on the inner boundaries of circumstellar disks can be obtained by exploring a more varied range of $M_{\mathrm{0}}$, $\alpha$, and $\beta_{\mathrm{rot}}$. This was not done due to computational constraints, as the simulations are expensive to run at such high resolution (each run used at least 100 000 CPU hours in total).

\subsection{Consequences on global disk evolution scenarios}

As a result of the high values of $n_{\mathrm{acc}}$ hereby reported, the mass of circumstellar disks in numerical simulations may have been underestimated by about an order of magnitude \citep{hennebelle_disks}. This appears to be inconsistent with observations of class 0 disks, which report lower disk masses of $\sim 10^{-2}\ \mathrm{M_{\odot}}$ \citep{tobin_2020}. We note however, that a recent study by \cite{duy_2024} has found that class 0 disk masses are routinely underestimated with current observational techniques. Nevertheless, the results indicate that the disk is very massive at birth, and naturally one must question how such a disk might evolve over time.
\\
\\
As such, we put forward two speculative evolutionary scenarios that we believe to be possible. The first scenario rests on the previously discussed fragmentation of the disk; as accretion subsides and the disk cools, it will become more prone to gravitational instabilities. Should the inner disk fragment, a significant amount of angular momentum could be extracted from the system. 
\\
Another scenario would rest on the strength of magnetic fields in the inner disk. Perhaps the results of current state of the art papers that report high thermal to magnetic pressure ratios are overestimating the strength of magnetic resistivities in the gas whose density exceeds first Larson core densities ($\rho > 10^{-13}\ \mathrm{g\ cm^{-3}}$), and thus would be underestimating the magnetic field strength within the protostar's and inner disk's precursor. This is in light of new studies (\citealp{lebreuilly_2023c, kawasaki_2023, tsukamoto_2023}; Vallucci-Goy et al. 2024 in-prep) that cast doubt on the validity of the MRN (Mathis-Rumple-Nordsiek, \citealp{mathis_1977}) dust size distribution used in said state of the art papers. In the case where the resistivities inside the first Larson core are overestimated, we would expect stronger coupling within the gas prior to hydrogen ionization. Although the ratio of thermal to magnetic pressure is still expected to greatly exceed unity within the inner disk, this would undoubtedly increase the strength of magnetic torques and cause a more central distribution of material, in which the protostar quickly exceeds the inner disk's mass and separates itself from it. However, simulations employing MRN resistivities create disk radii in broad agreement with observational surveys \citep{maury_2019, tobin_2020}, in which the magnetic field truncates the disk radius at $\sim 10^{1}\ \mathrm{AU}$. Perhaps the disk radii can serve as a lower bound on magnetic resistivities for low density gas, whereas the strength of magnetic fields in young stellar objects ($\sim 10^{3}\ \mathrm{G}$, \citealp{johns_2009}) can serve as an upper bound on resistivities within the first Larson core.

\subsection{Comparison with previous works}

\noindent We now turn to comparing our results with previous calculations in the literature that have resolved the birth of the protostar and the inner disk structure described in this paper. Due to the very stringent time-stepping, only a handful of such studies exist. To our knowledge, the first report of the existence of a swollen disk-like structure around the second core is that of \cite{Bate_1998}, who carried out their calculations using the SPH numerical method under the barotropic approximation. \cite{saigo_2008} later confirmed the existence of such a structure with their own barotropic calculations, this time using a nested-grid code. Although these papers do no have enough details presented for us to quantitatively compare our results, the description they provide of the structure of the second core and its surroundings seem to be qualitatively similar to ours. Following these two studies, \cite{machida_2011b} lead a detailed study of the inner disk under the barotropic approximation in which they include magnetic fields with ohmic dissipation, where they found qualitatively similar results to the hydro case. Notably, they report an inner disk mass of $\sim 10^{-2}\ \mathrm{M_{\odot}}$, and a protostellar mass of $\sim 10^{-3}\ \mathrm{M_{\odot}}$, in accordance with our results.
\\
More recent studies include those of \cite{wurster_2018, vaytet_2018, machida_2019, wurster_2020, wurster_2022}. The simulations presented in \cite{vaytet_2018} were also run with the {\ttfamily RAMSES} code while including magnetic fields. They reported the existence of a circumstellar disk around the protostar when including magnetic resistivities. They also report that they could not resolve the shock front separating the protostar from the inner disk, although as we have demonstrated in this paper, such a shock front does not exist. The mass of the disk reported in their paper is $\sim 10^{-4}\ \mathrm{M_{\odot}}$, which is to be expected given their short simulation time following protostellar birth ($24$ days) owing to stringent time-stepping.
\\
As stated previously, \cite{machida_2019} have also studied the inner disk, in which they were able to follow the calculations for a period of 2000 years following protostellar birth. Their results seem to indicate similar ratios of protostellar to disk mass than our paper, and the structure of their inner disk also seems to be similar to ours despite the presence of a magnetically launched outflow and a high velocity jet.
\\
Finally, \cite{wurster_2018, wurster_2020, wurster_2021, wurster_2022} ran these calculations using SPH and all non-ideal MHD effects: ohmic dissipation, ambipolar diffusion, and the hall effect. They found the same disk structure following the second collapse phase. In contrast to \cite{machida_2019}, they report no high velocity jets and argue that magnetically launched outflows are circumstantial and depend on the initial turbulent velocity vector field, as well as the non-ideal effects at play. The only way in which we may quantitatively compare our results to theirs is through the protostellar mass, which seems to be $\sim 10^{-3}\ \mathrm{M_{\odot}}$. The inner disk mass was not measured in these papers.
\\
In summary, the circumstellar disk structure described in this paper is routinely found in previous papers simulating the second collapse in 3D. Although the quantitative details of the properties of said disk may differ due to different numerical methods and physical setups, it is guaranteed to be recovered in simulations that possess sufficient amounts of angular momentum in the first core. As such, the only simulations that do not recover it are those that make use of the ideal MHD approximation, which extracts too much angular momentum from the system and prevents the protostar from ever reaching breakup velocity.

\section{Conclusion}
We have carried out a set of high resolution 3D RHD simulations that self-consistently model the collapse of a $1\ \mathrm{M_{\odot}}$ dense molecular cloud core to stellar densities with the goal of studying the innermost ($< 1\ \mathrm{AU}$) regions. Our results can be summarized as follows:
\begin{enumerate}[label=(\roman*)]
 \item Following the second gravitational collapse, the protostar is formed through hydrostatic balance. Through accretion, the protostar accumulates angular momentum and reaches breakup velocity, after which it sheds some of its mass to form a hot, dense, and highly turbulent circumstellar disk, which we call the inner disk. The protostar is embedded within this disk, and no shock front separates the two. As accretion continues, the disk completely engulfs the protostar and spreads outward due to a combination of excess angular momentum and accretion. The disk mass exceeds that of the protostar approximately seven fold, which means that the majority of the mass following the second collapse resides in the disk and its self-gravity dominates, with notable contributions from thermal pressure on its dynamics. In the case where an outer disk exists prior to the second collapse, this circumstellar disk forms within it, and the two merge after the inner disk spreads to sufficiently large radii.
 \item Despite the differing evolutionary histories at larger spatial scales, the star-disk structure formed after the onset of the second collapse is identical, with a small spread caused by the turbulent initial conditions. This is due to the formation of the first Larson core in all our simulations: a hydrostatic object that ensures that the second collapse occurs in approximately similar conditions.
 \item Accretion onto the protostar mainly occurs through material that slides on the disk's surface, as polar accretion has a low mass flux in comparison. Along the equator, material spreads outward due to excess angular momentum. Accretion onto the inner disk is highly anisotropic.
  \item The radiative emissions of the star-disk system is anisotropic: the radiative efficiency of the accretion shock is supercritical along the poles, whereas the inner disk's equatorial shock front is subcritical and of the order of $10^{-3}$.
  \item The density of the inner disk's shock front at $1\ \mathrm{AU}$ (the most commonly used sink radius) is in the range of $[5.35\times 10^{-10}; 2\times 10^{-9}]\ \mathrm{g\ cm^{-3}}$, which is about an order of magnitude higher than the commonly used sink accretion threshold of $1.66\times 10^{-11}\ \mathrm{g\ cm^{-3}}$. Thus, we suggest higher accretion thresholds for studies employing sink particles whenever possible. We note however, that our results correspond to very early times and may not be applicable throughout the entirety of the class 0 phase.
 \item In order to physically decouple the protostar from its disk and reduce its rotation rate to observed values, torque mechanisms need to transport a sufficient amount of angular momentum outward. Magnetic fields are likely to play this role.
\end{enumerate}
These results reveal the structure and kinematics of the innermost regions of circumstellar disks, which are often omitted from simulations due to computational constraints.
Although our results are valid for the timescales described here (< 6 years following protostellar birth), we expect magnetic fields to play a more significant role later on, particularly in creating powerful torques that transport material toward the protostar.

\begin{acknowledgements}
      This work has received funding from the French Agence Nationale de la Recherche (ANR) through the projects COSMHIC (ANR-20-CE31- 0009), DISKBUILD (ANR-20-CE49-0006), and PROMETHEE (ANR-22-CE31-0020). We have also received funding from the European Research Council synergy grant ECOGAL (Grant : 855130). We thank Ugo Lebreuilly, Anaëlle Maury, and Thierry Foglizzo for insightful discussions during the writing of this paper. The simulations were carried out on the Alfven super-computing cluster of the Commissariat à l'Énergie Atomique et aux énergies alternatives (CEA). Post-processing and data visualization was done using the open source \href{https://github.com/osyris-project/osyris}{Osyris} package. 3D visualizations were done using the open-source \href{https://docs.pyvista.org/version/stable/}{PyVista} package \citep{sullivan_2019}.
\end{acknowledgements}

\bibliographystyle{aa}
\bibliography{biblio}

\appendix

\section{Defining the protostar and inner disk in our simulations}
\label{appendix:definitions}
The definition of the protostar in our simulations is rather semantic, as it is not a separate object from the inner disk. We use the same criterion as \cite{vaytet_2018} for defining the protostar: it is simply the gas whose density is above $10^{-5}\ \mathrm{g\ cm^{-3}}$. We note that in \cite{vaytet_2018}, the authors mention that they do not have enough resolution to resolve the protostellar accretion shock separating it from the circumstellar disk, although as we have shown in \hyperref[section:breakup]{Sec. \ref*{section:breakup}}, such a discontinuity does not exist.
\\
\\
In order to define the inner disk, none of the criteria currently used in the literature were adequate in our case, mostly due to our use of turbulent initial conditions. Additionally, in some simulations, the second collapse occurred within a larger scale disk, which further complicated our disk selection criterion. Inspired by \cite{joos_2012}, we used the following criterion:

\begin{equation*}
      \begin{cases}
    P > \rho v_{r}^{2}, & \text{Thermal support against radial infall}\\
    \rho v_{\phi}^{2} > P, & \text{Centrifugal support exceeds pressure}\\
    \rho_{\mathrm{s}} < \rho < 10^{-5}\ \mathrm{g\ cm^{-3}}, & \text{Density threshold}
  \end{cases}
\end{equation*}
where $\rho_{\mathrm{s}}$ is the density of the inner disk's equatorial shock front, obtained at each snapshot through ray-tracing. This criterion ensures that no cells currently undergoing the second gravitational collapse phase are selected, and that sufficient angular momentum is present in the gas to qualify as a disk. The third item in this criterion states that $\rho < 10^{-5}\ \mathrm{g\ cm^{-3}}$, although this is again an arbitrary choice to separate the protostar from its circumstellar disk.


\end{document}